\PassOptionsToPackage{colorlinks=true, urlcolor=blue, linkcolor=blue, citecolor=blue}{hyperref}
\documentclass[%
 reprint,
 amsfonts,amsmath,amssymb,
 aps,
 prd
]{revtex4-2}

\RequirePackage{graphicx}
\usepackage{amsmath}
\usepackage{amsfonts}
\usepackage{amsthm}
\usepackage{mathrsfs}
\usepackage{amssymb}
\usepackage{epsfig}
\usepackage{color}
\usepackage{amscd}
\usepackage{xcolor}
\usepackage{nicefrac}
\usepackage{subfigure}
\usepackage{orcidlink}
\usepackage[%
  colorlinks=true,
  urlcolor=blue,
  linkcolor=blue,
  citecolor=blue
]{hyperref}

\def\prd{Phys. Rev. D }

\def\mnras{Monthly Notices of the Royal Astronomical Society }
\def\apj{The Astrophysical Journal }
\def\aap{Astronomy and Astrophysics }

\newcommand{\beq}{\begin{equation}}
\newcommand{\eeq}{\end{equation}}
\newcommand{\bea}{\begin{eqnarray}}
\newcommand{\eea}{\end{eqnarray}}

\providecommand{\dif}{\mathrm{d}} \def\d{\dif}

\def\af{\zeta}

\def\p{p}  \def\cp{\pi}
\def\mJ{\rm I} \def\mD{\rm II}

\def\mir{\mathrm{r}}
\def\mit{\mathrm{\theta}}
\def\mip{\mathrm{\phi}}

\newcommand{\aaa}{{\diamond}}

\newcommand{\ce}{{\cal{E}}} 
\newcommand{\cl}{{\cal{L}}} 
\newcommand{\cb}{{\cal{B}}} 

\newcommand{\dd}{d}
\newcommand{\di}{d_i}
\newcommand{\dI}{\bar{d}_i}
\renewcommand{\dj}{d_j}
\newcommand{\dJ}{\bar{d}_j}
\newcommand{\ei}{e_i}
\newcommand{\eI}{\bar{e}_i}
\newcommand{\ej}{e_j}
\newcommand{\eJ}{\bar{e}_j}
\newcommand{\elE}{\mathrm{E}}
\newcommand{\elK}{\mathrm{K}}
\newcommand{\elPi}{\Pi}
\newcommand{\rp}{r_{\rm p}}
\newcommand{\rpi}{r_i}
\newcommand{\rpI}{\bar{r}_i}
\newcommand{\lp}{l_s}
\newcommand{\lm}{l_c}
\def\Opava{Research Centre for Theoretical Physics and Astrophysics, Institute of Physics, Silesian University in Opava, CZ-74601 Opava, Czech Republic}
\def\Praha{Institute of Theoretical Physics,\\ Faculty of Mathematics and Physics,
Charles University,\\
V~Hole\v{s}ovi\v{c}k\'{a}ch 2, CZ-180\,00 Prague 8, Czech Republic}

\begin{document}

\title{
Charged particle dynamics in magnetosphere generated by current loop around Schwarzschild black hole
}

\author{Martin Kolo\v{s}\; \orcidlink{0000-0002-4900-5537}}\email{martin.kolos@physics.slu.cz}\affiliation{\Opava}
\author{David Kofro\v{n}\;\orcidlink{0000-0002-0278-7009}}\email{d.kofron@gmail.com}\affiliation{\Opava\vspace{1ex}\\ \Praha}

\begin{abstract}
We present a theoretical study of the magnetic field generated by a toroidal current loop situated in the equatorial plane of a non-rotating Schwarzschild black hole, based on the dynamics of charged particles. Using the exact general relativistic solution for the magnetic field, we analyze particle motion both analytically and numerically, identifying regions of stable and unstable orbits. In particular, we classify charged particle dynamics into attractive and repulsive Lorentz force configurations and show that in the attractive case, charged particles can accumulate near the current loop, forming collective currents that oppose the original current loop magnetic field. We demonstrate that charged particle accumulation can lead to the formation of toroidal structures analogous to radiation belts in the BH magnetosphere.
We compare the curved spacetime solution to flat spacetime analogs and highlight general relativistic effects such as the existence of the innermost stable circular orbit (ISCO) for charged particles, which sets a lower bound for radiation belt formation. The divergence of the vector potential at the loop location in the idealized infinitesimal loop model is addressed, and we argue that a physically realistic model must consider a finite-width current distribution to avoid unphysical divergences in the effective potential. 
\end{abstract}


\maketitle


\section{Introduction}

Long-range gravitational and electromagnetic interactions play a key role in understanding high-energy processes near black holes (BHs). Observational evidence strongly indicates the presence of magnetic fields in the vicinity of astrophysical BHs \cite{Daly:2019:APJ:}. Orders of magnitude of magnetic fields around BHs may vary from few Gauss up to $10^8$Gauss, depending on the source generating the field. For stellar-mass BHs observed in X-ray binaries the characteristic strength of magnetic fields are of order $10^8$~G, while for supermassive BHs the characteristic strength is of order $10^4$~G. Since the energy densities of magnetic fields of such orders are not enough to make sufficient contribution to the geometry of the background spacetime, in realistic astrophysical situations the spacetime metric around a BH can be fully described by the Kerr solution of the Einstein field equations parameterized by the mass and the angular momentum of the BH. The structures of magnetic fields around astrophysical BHs have not yet been properly resolved. In some regions with relatively higher mass density, the magnetic field is expected to have a quite complex character due to turbulent processes inside the surrounding plasma. Yet, in predominantly dilute regions, like those outside an accretion disk, the field lines can be of regular and even extended large-scale shape, which is supported by polarimetric observational studies related to various astrophysical BH candidates and their relativistic jets \cite{Nak-etal:2018:APJ:}.

For the simplest approximation of the BH magnetosphere problem, one can begin with the vacuum solutions of the Maxwell equations in a curved background. Given that the Maxwell equations are linear, it is possible to combine two or more solutions to generate a new one. Here, we use a current loop located in the equatorial plane of a nonrotating Schwarzschild BH as the magnetic field source, closely related to studies by Petterson \cite{Petterson:1975:PRD:,Chi-Vis:1975:PRD:,Moss:2011:PRD:}. There are exact solutions to Maxwell’s equations in curved backgrounds, such as Wald’s solution for a uniform magnetic field \cite{Wald:1974:PHYSR4:} or the Blandford and Znajek solution \cite{Bla-Zna:1977:MNRAS:}, commonly applied in BH jet models. The precise structure of the magnetosphere around astrophysical BHs is strongly influenced by accretion processes, which are complex and nonlinear; hence, analytical models are inadequate to capture all turbulent nonlinear effects. Numerical simulations, like relativistic magnetohydrodynamics (GRMHD) \cite{Tch-Nar-McK:2010:APJ:,Kol-Jan:2020:RAG:,Jan-Jam:2022:AAP:} or relativistic particle-in-cell (GRPIC) simulations \cite{Par-Phi-Cer:2019:PRL:,Cri-etal:2021:AAP:}, are essential in such cases. As we can see, the modeling of realistic magnetic field lines in a strong gravity regime remains challenging, and in this contribution we will explore model for BH magnetosphere generated by single current loop located in BH equatorial plane. As we will see, even it this elementary analytic model the dynamic of charged test particles can be very complex.

\begin{figure}
\centering
\includegraphics[width=0.4\textwidth]{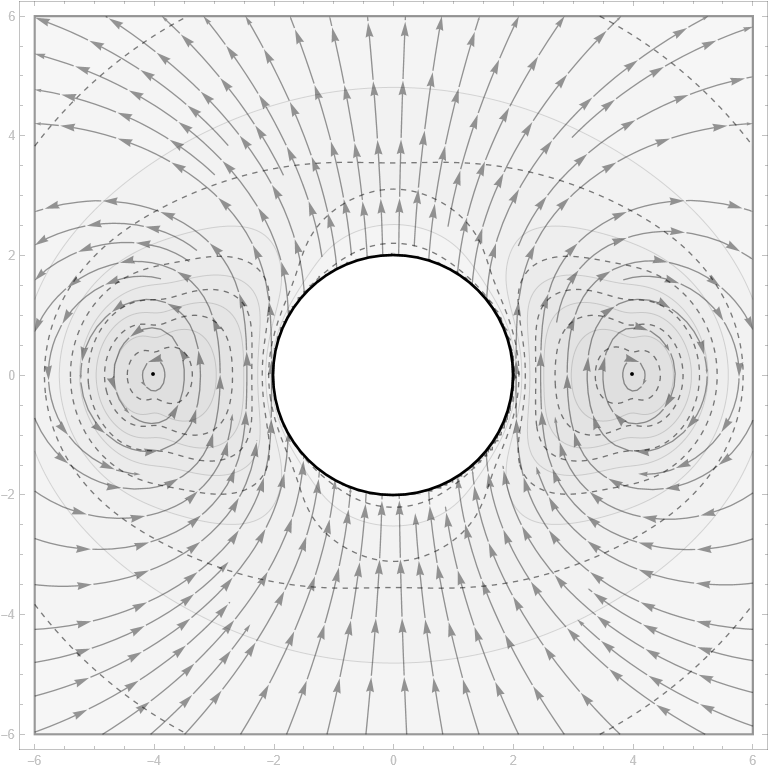}
\caption{Magnetic field lines generated by an electric current loop in the equatorial plane of Schwarzschild BH. The contours and appropriate densities depict the magnitude of the field. The values of the constants are $\rp=2\,,r_0=4$ (yet, the absolut scale is irrelevant).}
\label{figMFa}
\end{figure}

Magnetic fields in strong gravity regions are vital for explaining various high-energy phenomena observed in BH systems across different mass scales. Such phenomena include, for example, the formation and collimation of relativistic jets \cite{Bla-Zna:1977:MNRAS:} in active galactic nuclei (AGNs) and other BH systems, the generation of high-frequency quasi-periodic oscillations (QPOs) in microquasars and AGNs \cite{Kol-Stu-Tur:2015:CLAQG:}, and the acceleration of ultra-high-energy cosmic rays \cite{Tur-etal:2019:ApJ:}, among others. The Lorentz force acting on the charged particles can be substantial, or even dominant, especially for elementary particles. For a relativistic particle with charge $q$ and mass $m$, the ratio of the Lorentz force to the gravitational force near the BH can be represented by the following dimensionless parameter \cite{Fro-Sho:2010:PHYSR4:}
\beq 
{\cal B} =\frac{|q| G M B}{2 m c^4}. \label{BB-param}
\eeq 
This parameter can be significant even for weak magnetic fields due to the large value of the specific charge $q/m$ \cite{Fro-Sho:2010:PHYSR4:}. 

We can use Earth's magnetosphere observations as motivation for analysis of charged particle dynamic in the combined gravitational and electromagnetic field of the BH. The motion of a single charged particle is used to understand the fundamental mechanisms of trapping, drift, and stability in Earth’s magnetosphere, forming the basis for describing the collective plasma behavior. The charged particles can accumulate in the Van Allen radiation belts—zones of energetic electrons and ions held in place by Earth's magnetic field. The movement of accumulated particles, especially the drifting ions, generates a westward-flowing ring current around the equator at the altitude of several Earth radii. This ring current produces its own magnetic field, which opposes and thus weakens Earth's main magnetic field \cite{Kos-Kil:2022:BOOK:}. We are interested in whether analogous behavior can occur in the vicinity of BHs and whether it is possible to construct radiation belts in BH environment. 

In astrophysically realistic scenarios, the electromagnetic interaction parameter (for elementary particles) is typically so large that the Lorentz force vastly dominates over gravity. This allows for a relatively straightforward local description using plasma physics in the background of the curved spacetime. A particularly intriguing and novel situation arises when the gravitational and electromagnetic forces are of comparable magnitude. In such regimes, the full complexity of chaotic particle dynamics emerges, especially in regions near the BH horizon specifically, below the photon orbit in the case of a non-rotating Schwarzschild BH, or beneath the ergosphere in the case of a rotating Kerr BH. In these extreme environments, one may expect new and hitherto unexplored effects, such as the radiative Penrose process \cite{Kol-Tur-Stu:2021:PRD:}. 

The article is separated into three sections; in the first section we will provide the analytic solution of the Maxwel equations given by current loop and explore structure of generated magnetic field line; in the second section we investigate charged particle dynamic in thin BH magnetosphere model; in the last section we try to address the question of stability of a combined relation magnetic field particle interaction.

\section{Magnetic field generated by current loop} \label{sec:1}

We consider a BH of a mass $M$ described by the Schwarzschild metric 
\beq
\d s^2 = -f(r) \, \d t^2 + f(r)^{-1} \, \d r^2 + r^2( d \theta^2 + \sin^2\theta \, \d \phi^2), \label{SCHmetric}
\eeq
where $f(r)$ is the lapse function  
\beq 
 f(r) = 1 - \frac{2 M}{r}=\frac{\Delta(r)}{r^2}.
 \label{fun_f}
\eeq
We will be interested in the spacetime above the BH horizon limit $r>\rp=2M$.

The Maxwell equations expressed in Schwarzschild spacetime for the four-potential $A_\alpha=(0,0,0,A_\phi)$ are reduced to a single partial differential equation 
\beq \label{eq:Maxwell}
    \frac{1}{\sin\theta} \frac{\partial}{\partial r} \left( f(r)\frac{\partial A_\phi}{\partial r}\right)+
    \frac{1}{r^2}\frac{\partial}{\partial\theta}\left(\frac{1}{\sin\theta}\frac{\partial A_\phi}{\partial \theta}\right)=0.
\eeq
We will not solve the Maxwell equations directly. Instead, we will employ the Debye potential approach as in \cite{Kof-Kot:2022:PRD:}. The orthonormal components of the magnetic field vector $ \textbf{B} = (B_{\hat{r}},B_{\hat{\theta}},B_{\hat{\varphi}})$, observed by a congruence of static observers, are given by
\beq
    B_{\hat{j}} = 
    e_{\hat{j}}^{\phantom{\hat{i}}b}
    (\star F)_{ab} u^a, 
\eeq
where $(\star F)_{ab}$ is the Hodge dual of the Maxwell tensor $F_{ab}$ defined as $(\star F)_{ab}=\frac{1}{2}\,\epsilon_{abcd}\,F^{cd}$ with the help of volume 4-form $\epsilon_{abcd}$.

The orthonormal components of the magnetic field take the form 
\bea 
B_{\hat{r}} &=&  -\frac{1}{r^2\sin\theta}\frac{\partial A_\phi}{\partial \theta}\,, \\
B_{\hat{\theta}} &=& \sqrt{f(r)}\,\frac{1}{r\sin\theta}\frac{\partial A_\phi}{\partial r} \,, \\
B_{\hat{\phi}} &=& 0\,,
\eea
in terms of the 4-potential potential
\beq
    \boldsymbol{B}=\frac{1}{r^2\sin\theta}\left(
    -\frac{\partial A_\phi}{\partial \theta}\,\boldsymbol{e_r}+
    r\sqrt{f(r)}\,\frac{\partial A_\phi}{\partial r}\,\boldsymbol{e_\theta}\right),
\eeq 
for $ \boldsymbol{A} = A_\phi\, \boldsymbol{\mathrm{d}\phi} $. The square of magnetic field strength $\textbf{B}$ as a function of $r$ and $\theta$ measured by observers takes the form
\beq
\textbf{B}^2 = B_{\hat{r}}^2+B_{\hat{\theta}}^2 , \label{magnitude}
\eeq

\begin{figure*}
\centering
\includegraphics[width=\textwidth]{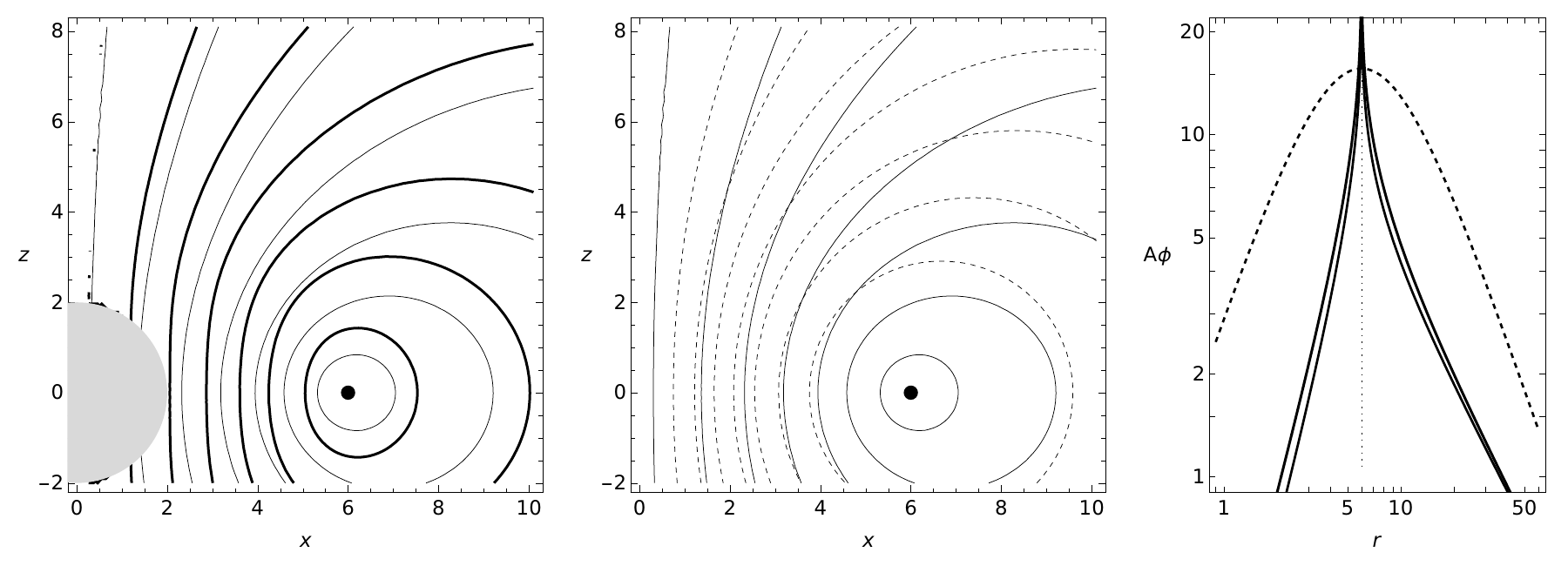}
\caption{ 
Contour of density for electromagnetic four-potential $A_{\phi}(r,\theta)$ giving the shape of magnetic field generated by equatorial current loop in BH spacetime. 
In the left figure we show the magnetic field lines for full analytic solution $A_{\phi}$ in curved BH spacetime (thick curves) comparing it to its flat spacetime analog $A_{\phi}^{\rm F}$ (thin). The same values for both $A_{\phi}$ are used; the differences between flat and BH solutions will became smaller when the current loop will be moved far away from the central BH.
In the middle figure we compare flat spacetime analytic solution $A_{\phi}^{\rm F}$ (solid curves) with simple heuristic model $A_{\phi}^{\rm H}$ given by (\ref{power1st}).
In the right figure we plot electromagnetic four-potential as function of radius $A_{\phi}(r)$ for all three cases (curved, flat, heuristic) demonstrating the divergence of both full analytic solutions (flat,curved) at the current loop radius (dotted line).
\label{figMF}
}
\end{figure*}

\subsection{Full analytic solution}

Let us recall the textbook result: the electromagnetic four-potential $A_\alpha=(0,0,0,A_\phi)$ of a current loop of radius $a$ in flat spacetime (expressed in spherical coordinates and for with unit current $\mu I =1$ ) is given by \cite{Jackson:1998:book:}
\beq 
A^{\rm F}_\phi(r,\theta)=\frac{1}{\pi}\, \sqrt{d_{\rm F}}\, \Bigl[ (2-m_{\rm F})\elK(m_{\rm F})-2\elE(m_{\rm F}) \Bigr]\,,
\label{eq:Aflat}
\eeq
where we denoted
\begin{align}
    d_{\rm F} &= a^2+2ar\sin\theta+r^2 \,,\\
    m_{\rm F} &= \frac{4ar\sin\theta}{d_{\rm F}} \,,
\end{align}
and the super/sub-script $\cdot_{\rm F}$ refers to flat spacetime.

The complete analytic solution in a closed form for the electromagnetic field generated by a current loop around the rotating Kerr BH was given in \cite{Kof-Kot:2022:PRD:}. In this article, we will focus on the nonrotating Schwarzschild BH case, where the magnetic field generated by the current loop with unit current ($\mu I =1$) in the equatorial plane is given by four-potential
\begin{widetext} 
\begin{align}
\pi \, A_\phi (r,\theta) &= 
-2\sqrt{\dd} \, \elE\left(m\right)
-\frac{1}{\sqrt{\dd}}\left[\left(\dI\dJ+\di\dj\right)\frac{2r_0+\rp}{2r_0-\rp}-4r r_0\frac{2r-\rp}{2r_0-\rp}\right]\elK\left(m\right) \nonumber\\
&\phantom{=}-\frac{\rp \cos\theta}{\sqrt{\dd}} \Bigl[\left(\ej+\rpI\right)\elPi\left(n,m\right)+\left(\eJ+\rpi\right)\elPi\left(\bar{n},m\right)\Bigr]\nonumber\\
&\phantom{=}-\rp\,\frac{2r_0-\rp}{2\sqrt{\dd}}
\Biggl[
 \lp\, \frac{\ei-\rpi}{\ei\rpi} \, \elPi\left(+\frac{\rpi}{\ei}\,n,m\right)
+\lm\, \frac{\eJ+\rpi}{\eJ\rpi} \, \elPi\left(-\frac{\rpi}{\eJ}\,\bar{n},m\right) \nonumber\\
&\phantom{-\rp\,\frac{2r_0-\rp}{4\sqrt{\dd}}}
\;+\lp\, \frac{\eI-\rpI}{\eI\rpI} \, \elPi\left(+\frac{\rpI}{\eI}\,\bar{n},m\right)
  +\lm\, \frac{\ej+\rpI}{\ej\rpI} \, \elPi\left(-\frac{\rpI}{\eJ}\,n,m\right)
\Biggr] \nonumber\\
&\phantom{=}-2\pi \rp\Biggl[+
\cos^2\left(\nicefrac{\theta}{2}\right) \Theta(\cos\theta) 
\Bigl(-1+\Theta\Bigl(2(\di+\dI)(\rp^2+4r_0\rp-4r_0^2)-4\rp(\di\dj+\dI\dJ)\Bigr) \nonumber\\
&\phantom{=-2\pi \rp\Biggl[}
+\sin^2\left(\nicefrac{\theta}{2}\right) \Bigl(-1+\Theta(\cos\theta) \Bigr)
\Theta\Bigl(2(\di+\dI)(\rp^2+4r_0\rp-4r_0^2)+4\rp(\di\dj+\dI\dJ)\Bigr)+1 \Biggr] \label{eq:Aphi}
\end{align}
\end{widetext}
with (neither $i$ nor $j$ are summation indices)
\begin{align}
    \lp &= \left(r-\rp\sin^2(\nicefrac{\theta}{2})\right)\sin^2(\nicefrac{\theta}{2})\\
    \lm &= \left(r-\rp\cos^2(\nicefrac{\theta}{2})\right)\cos^2(\nicefrac{\theta}{2})\\
    \ei &= \frac{1}{2}\left(2r-\rp\right)\cos\theta+\frac{\rp}{2}-i\sqrt{r(r-\rp)}\,\sin\theta\,,\\
    \ej &= \frac{1}{2}\left(2r-\rp\right)\cos\theta-\frac{\rp}{2}-i\sqrt{r(r-\rp)}\,\sin\theta\,,\\
    \rpi &=\frac{\rp}{2}+i\sqrt{r_0(r_0-\rp)}\,,
    \end{align}
and
\bea
& \di = \eI-\rpi \,, \quad
\dj = \ej+\rpI \,, \quad
\dd = \di \dI \,, \\
& m = \frac{\di\dI-\dj\dJ}{\di\dI}\,,\quad
n =\frac{\dJ-\dI}{\di}.
\eea
This solution has two parameters: $\rp=2M$ is the position of the BH horizon and $r_0$ is the position of the current loop.

The rather surprising presence of Heaviside's step function $\Theta$ in the 4-potential is a consequence of the existence of a branch cut of the elliptic integrals of the third kind $\elPi(n,m)$. There is one in the complex plane that lies on the real line $n\in \langle 1,\infty)$. In the end, the $\Theta$ function counterbalances these jumps and the resulting potential is smooth. For detailed explanation of the complications caused by $\elPi(n,m)$ see the discussion \cite{Kof-Kot:2022:PRD:}. 

In further calculations, we will employ the limits of the function $A_\phi$ and its derivatives in the equatorial plane. They can be easily computed, yet less easily simplified, to the following
\begin{widetext}
\begin{align}
    A_\phi(r,\nicefrac{\pi}{2}) &= \pi \rp \Theta(r-r_0)-2\left(\sqrt{\Delta}+\sqrt{\Delta_0}\right) \elE(m)+
    2\,\frac{r^2(2r_0-\rp)+r \rp^2+r_0(r_0-\rp)(2r_0+\rp)}{(2r_0-\rp)\left(\sqrt{\Delta}+\sqrt{\Delta_0}\right)}\,\elK(m) \nonumber\\
    &\phantom{=} +i
    \frac{\sqrt{\Delta}-\sqrt{\Delta_0}}{\sqrt{\Delta}+\sqrt{\Delta_0}} \,
    \Delta'\Delta'_0\,\rp \left[
    \frac{1}{\left(2\sqrt{\Delta}-i\rp\right)\left(2\sqrt{\Delta_0}-i\rp\right)}\,
    \elPi\left(\frac{2\sqrt{\Delta}}{\sqrt{\Delta}+\sqrt{\Delta_0}} \frac{2\sqrt{\Delta_0}-i\rp}{2\sqrt{\Delta}-i\rp}\,,m\right)-\mathrm{c.c.}\right], \\
    \frac{\partial A_\phi(r,\nicefrac{\pi}{2})}{\partial \theta} &=0 \,, \label{eq:dAphi}\\
    \frac{\partial^2 A_\phi(r,\nicefrac{\pi}{2})}{\partial \theta^2} &=
    \frac{\sqrt{\Delta}+\sqrt{\Delta_0}}{(r-r_0)^2(r+r_0-\rp)^2}\,\left[r^3\rp+r_0^2(r_0-\rp)\rp+r r_0\rp(r_0+\rp)-r^2(4r_0^2-r_0\rp+\rp^2)\right] \elE(m) \nonumber\\
    &\phantom{=}-\frac{(r+r_0)\rp}{\sqrt{\Delta}+\sqrt{\Delta_0}}\,\elK(m) \,,\\
    \frac{\partial A_\phi(r,\nicefrac{\pi}{2})}{\partial r} &= 
    -\frac{2\Delta+r_0\rp}{(r-\rp)\left(\sqrt{\Delta}-\sqrt{\Delta_0}\right)}\,\elE(m) +
    \frac{2\Delta-r_0\rp}{(r-\rp)\left(\sqrt{\Delta}+\sqrt{\Delta_0}\right)}\,\elK(m)\,,\\
    \frac{\partial^2 A_\phi(r,\nicefrac{\pi}{2})}{\partial r^2} &=\frac{d}{d r}\left(\frac{\partial A_\phi(r,\nicefrac{\pi}{2})}{\partial r}\right) \label{eq:dArr}
\end{align}
\end{widetext}
where the modulus  $m$ is easily evaluable in the equatorial plane and $\Delta=\Delta(r),\,\Delta_0=\Delta(r_0)$ are merely shortcuts and "c.c." means complex conjugate of the previous terms in parentheses.

%

\subsection{Special limits of full solution}

First, as a consistency check, explicit calculations show that the flat space-time limit is recovered, that is, 
\beq 
\lim_{\rp\rightarrow 0} A_\phi(r,\theta) = A^{\rm F}_\phi(r,\theta).
\eeq
At infinity, the four-potential $A_\phi(r,\theta)$ (\ref{eq:Aphi}) is constant, and the additive constant has been chosen so that it equals zero and thus coincides with the flat spacetime solution $A^{\rm F}_\phi(r,\theta)$ at infinity. There is a logarithmic divergence in the position of the current itself $r=r_0$ such that the Amp\`ere\,--\,Maxwell law holds.

\subsubsection{Uniform magnetic field}

In the limit $r_0\rightarrow\infty$ (while increasing the current with the distance), we get a homogeneous magnetic field. The four-vector potential for the uniform magnetic field with an asymptotic value of the strength $B$ and the field lines orthogonal to the BH's equatorial plane in the Schwarzschild metric has the only non-zero component 
\beq
A_{\phi}^{\rm U} = B_{\rm U} \, r^2 \sin^2 \theta,\label{Auniform}
\eeq
where $B_{\rm U}$ is magnetic field magnitude. The uniform magnetic field around the BH can be interpreted as an external large-scale magnetic field whose source is located outside and far away from the BH.  Any source of the magnetic field located far away from BH will generate an approximately homogeneous uniform magnetic field, at least locally.

\subsubsection{Dipole magnetic field}

In contrast to the previous case, it is not possible to proceed with the limit $r_0\rightarrow 0$, for the four-potential (\ref{eq:Aphi}) of the current loop, as already discussed in \cite{Punsly2009}, since it is forbidden to lower the current loop under the BH horizon. One could na\"ively expect to recover a magnetic field of the magnetic dipole 
\beq
 A_{\phi}^{\rm D} = B_{\rm D} \left[ \ln\left( 1 - \frac{r_{\rm p}}{r} \right) + \frac{r_{\rm p}}{r} \left( 1+ \frac{r_{\rm p}}{2r} \right) \right] r^2 \sin^2\theta, \label{eq:dipole}
\eeq
but it is \emph{not} the case.

\begin{figure*}
\centering
\includegraphics[width=\textwidth]{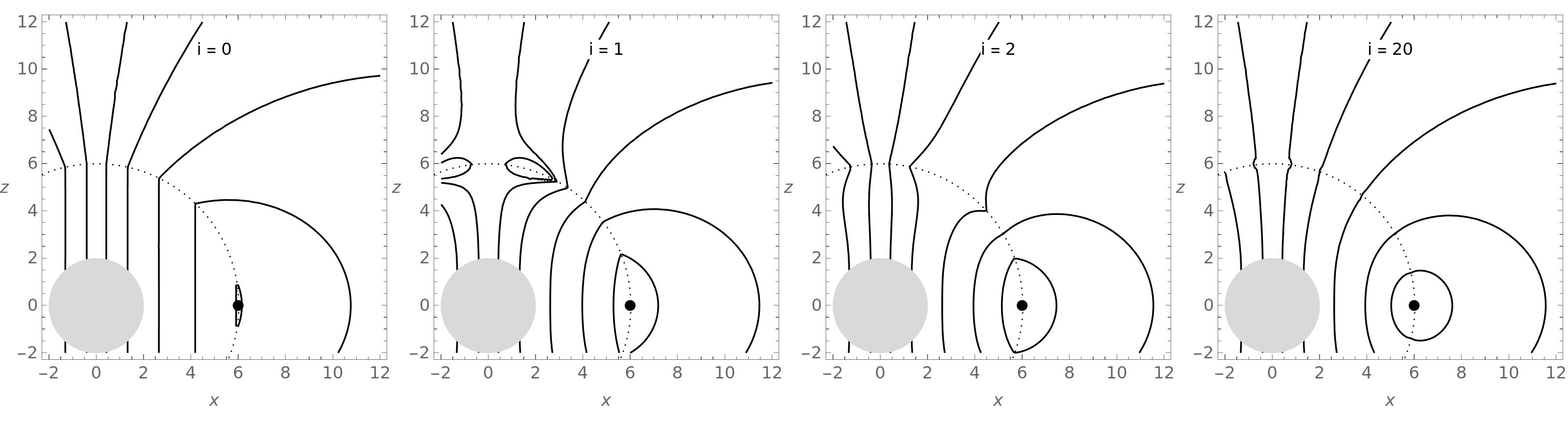}
\caption{Multipole expansion of the magnetic field generated by current loop (black dot) in equatorial plane as obtained by \cite{Petterson:1975:PRD:}. Discontinuity between inner and outer solution located close to sphere (dotted curve) of with the same radius as the current loop is clearly visible for first terms of this expansion, but will prevail for higher terms as well. \label{fig:expMulti}
}
\end{figure*}

\subsection{Expansion into infinite series}

In \cite{Petterson:1975:PRD:}, Petterson derived a solution for the magnetic field produced by a stationary axisymmetric toroidal current loop in the equatorial plane of a Schwarzschild black hole using a multipole expansion of the vector potential. The expansion is based on spherical harmonics and Legendre functions, which allows the magnetic field to be expressed as a superposition of individual multipole moments
\beq
A_{\phi}^{\rm E} = \sum_{l=0,2}^{\infty} r_{\rm p}^l  \, {\rm M}(l) \, R_l(r) \, C_l^{3/2}(\cos \theta)\, \sin^2 \theta, \label{eq:petterson}
\eeq
where $\rm{C}^{3/2}_l(\cos\theta)$ are Gegenbauer polynomials (special case of the Jacobi polynomials). We sum only terms with $l$ even because of refection symmetry about the equatorial plane. The special functions and coefficients are defined as
\beq
R_l(r) =
\begin{cases}
r^2 P_l^{(2,0)}\!\left(1 - \frac{2r}{r_{\rm p}}\right) V_l(r_0), & r_{\rm p} \leq r \leq r_0, \\[1ex]
{r_0}^2 P_l^{(2,0)}\!\left(1 - \frac{2r_0}{r_{\rm p}}\right) V_l(r), & r_0 \leq r,
\end{cases}
\eeq
where $P_l^{(2,0)}$ are Jacobi polynomials and
\bea
{\rm M}(l) &=& - \mathrm{i}^l \, (2 l+3) \sqrt{\frac{3 \pi }{2}} \sqrt{1-\frac{r_{\rm p}}{r_0}} \frac{ \Gamma \left(\frac{l+1}{2}\right)}{\Gamma \left(\frac{l}{2}+2\right)},\\ 
V_{l}(u) &=& - U_{l}(u) \int \frac{u \, \d u}{(1 - u) U_{l}^{2}(u)}, \\
U_{l}(u) &=& u^2 r_{\rm p}^{l+2} \, P_l^{(2,0)(1-2u) },
\eea
with $\Gamma(n)$ as Euler gamma function. The multipole expansion of $A_{\phi}^{\rm E}$ consists of two parts: the inner region ($r < r_0$) and the outer region ($r > r_0$). Only even terms contribute to the sum, while all odd terms vanish due to symmetry. An approximate solution (first term of the multipole expansion) valid outside the loop ($r > r_0$), where the series converges rapidly, is dominated by the dipole term (\ref{eq:dipole}). An approximate solution valid inside the loop ($r_{\rm p} < r < r_0$) corresponds to a uniform magnetic field (\ref{Auniform}).

Although each individual term in the infinite sum~(\ref{eq:petterson}) satisfies the Maxwell equations~(\ref{eq:Maxwell}), the matching of the inner and outer solutions is not smooth term by term. Only when the complete sum is considered does the discontinuity at the radius sphere $r = r_0$ disappear. Magnetic fields for different terms of the multipole expansion (\ref{eq:petterson}) are shown in Fig.~\ref{fig:expMulti}, where one can clearly observe how the discontinuity at $r = r_0$ diminishes with increasing expansion order, eventually converging to the full analytic solution. The existence of a discontinuity at $r = r_0$ poses a challenge for charged particle dynamics, as the particle’s kinematic momentum can change abruptly when crossing the $r = r_0$ shell. Despite these limitations, the Petterson multipole expansion remains a valuable analytical tool for studying large-scale magnetic field configurations in curved spacetime. The outer dipole solution can serve as a simplified general relativistic model for a neutron star magnetosphere (with the current loop located inside the star \cite{Vrb-Kol-Stu:2024:EPJC:}), while the inner uniform magnetic field (Wald's solution \cite{Wald:1974:PHYSR4:}) represents an useful approximation for the BH magnetosphere (with the current loop located at infinity).

An alternative expansion of the complete analytic solution in flat spacetime (\ref{eq:Aflat}) was mentioned in \cite{Jackson:1998:book:,Kolos:RAGtime:2017:} using powers series expansion in the modulus $m_{\rm F}$. The first therm of this expansion 
\beq \label{power1st}
A_{\phi}^{\rm H} (r,\theta) = 
B_{\rm H} \frac{r_0^2 r \sin\theta}{\left({r_0}^2+r^2\right)^{3/2}} + \ldots
\eeq
can be used as a heuristic model that approximates the magnetic field around the current loop with a simple sooth analytic formula. As we can see in Fig.~\ref{figMF}, the $A_{\phi}^{\rm H}$ (\ref{power1st}) solution is shifted from the full analytic formula $A_{\phi}^{\rm F}$ (\ref{eq:Aflat}), but the overall structure of the magnetic field lines is conserved. For $A_{\phi}^{\rm H}$ there is no discontinuity at the sphere of radius $r = r_0$ and charged particles can move through without sudden momenta change. The complete analytic solution (\ref{eq:Aphi}) is generated by the delta current flowing in an infinitely thin "wire" completely located at the radius $r_0$. The four-potential $A_{\phi}^{\rm H}$ (\ref{power1st}) is not a solution of the vacuum Maxwell equations, and smooth current density flowing in the toroidal direction is necessary for support; see the right subfigure in Fig.~\ref{figMF}.



\begin{figure*}
\centering
\includegraphics[width=0.45\textwidth]{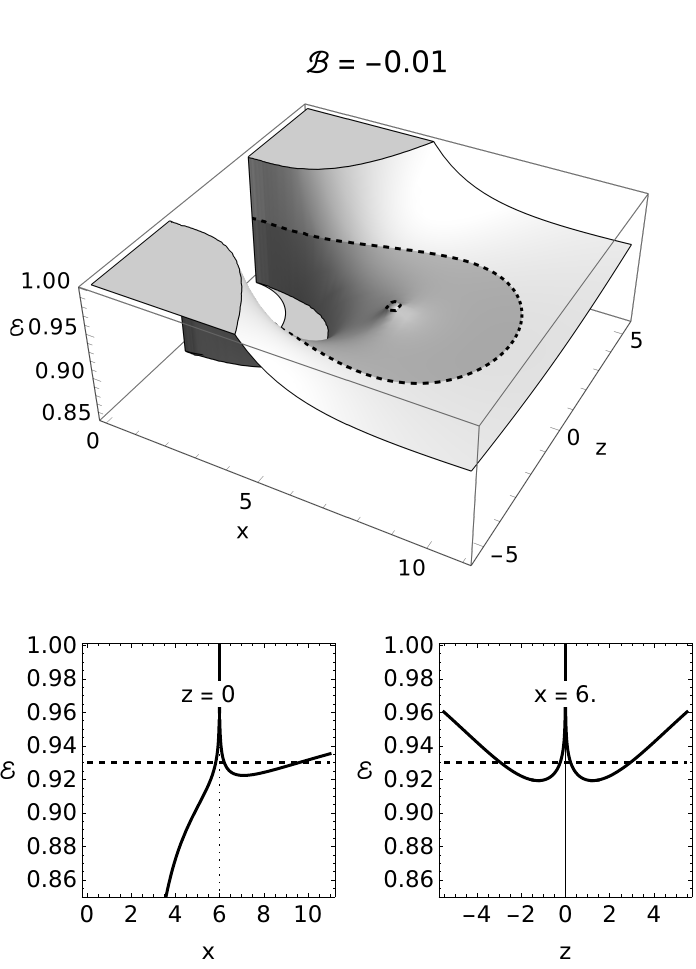}
\includegraphics[width=0.45\textwidth]{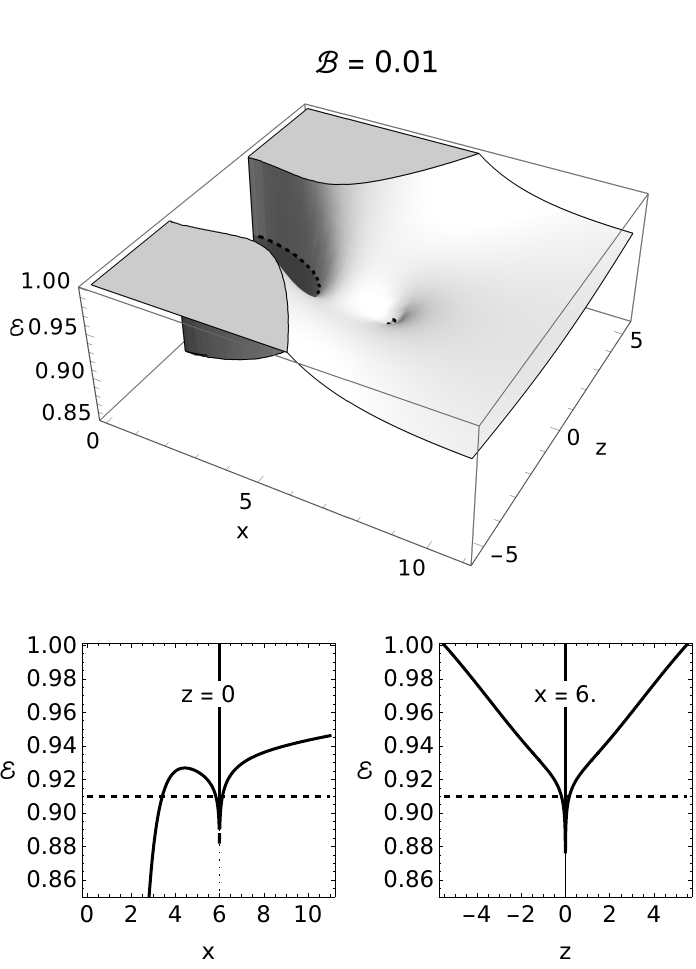}
\caption{Two examples of the effective potential as a function of coordinates $x,z$ for different $\cb$ parameters. We see a peak in the effective potential $V_{\rm eff}(x,z)$ at the current loop radius for $\cb=-0.01$ (repulsive Lorentz force from the current loop) while we see dip in $V_{\rm eff}$ for $\cb=0.01$ (attractive Lorentz force). Lets note, that $V_{\rm eff}(x,z)$ function diverges at current loop for both $\cb<0$ and $\cb>0$ cases. At lower sub-figures, cross section of the whole $V_{\rm eff}(x,z)$ function are plotted.
}
\label{fig:Veff}
\end{figure*}

\section{Charged particle dynamic in BH magnetosphere.}

\subsection{Equations of motion}

The motion  of charged test particle is described by the covariant Lorentz equation
\beq
	m \frac{D u^\mu}{\d \tau} = q {F^{\mu}}_{\nu} u^{\nu},
\eeq
where $u^{\mu}$ is the four-velocity of the particle with the mass $m$ and charge $q$, normalized by the condition $u^{\mu} u_{\mu} = -1$, $\tau$ is the proper time of the particle, and $F_{\mu \nu} = A_{\nu,\mu} - A_{\mu,\nu}$ is the antisymmetric tensor of the electromagnetic field. Using Hamiltonian formalism for the charged particle motion, we can write
\beq
  H =  \frac{1}{2} g^{\alpha\beta} (\cp_\alpha - q A_\alpha)(\cp_\beta - q A_\beta) + \frac{1}{2} \, m^2
  \label{particleHAM},
\eeq
where the kinematical four-momentum $\p^\mu = m u^\mu$ is related to the generalized (canonical) four-momentum $\cp^\mu$ by the relation
\beq
 \cp^\mu = \p^\mu + q A^\mu, \label{particleMOM}
\eeq
that satisfies the Hamilton equations in the form
\beq
 \frac{\d x^\mu}{\d \af} \equiv \p^\mu = \frac{\partial H}{\partial \cp_\mu}, \quad
 \frac{\d \cp_\mu}{\d \af} = - \frac{\partial H}{\partial x^\mu}. \label{Ham_eq}
\eeq
The affine parameter $\af$ of the particle is related to its proper time $\tau$ by the relation $\af=\tau/m$.

\begin{figure*}
\centering
\includegraphics[width=\textwidth]{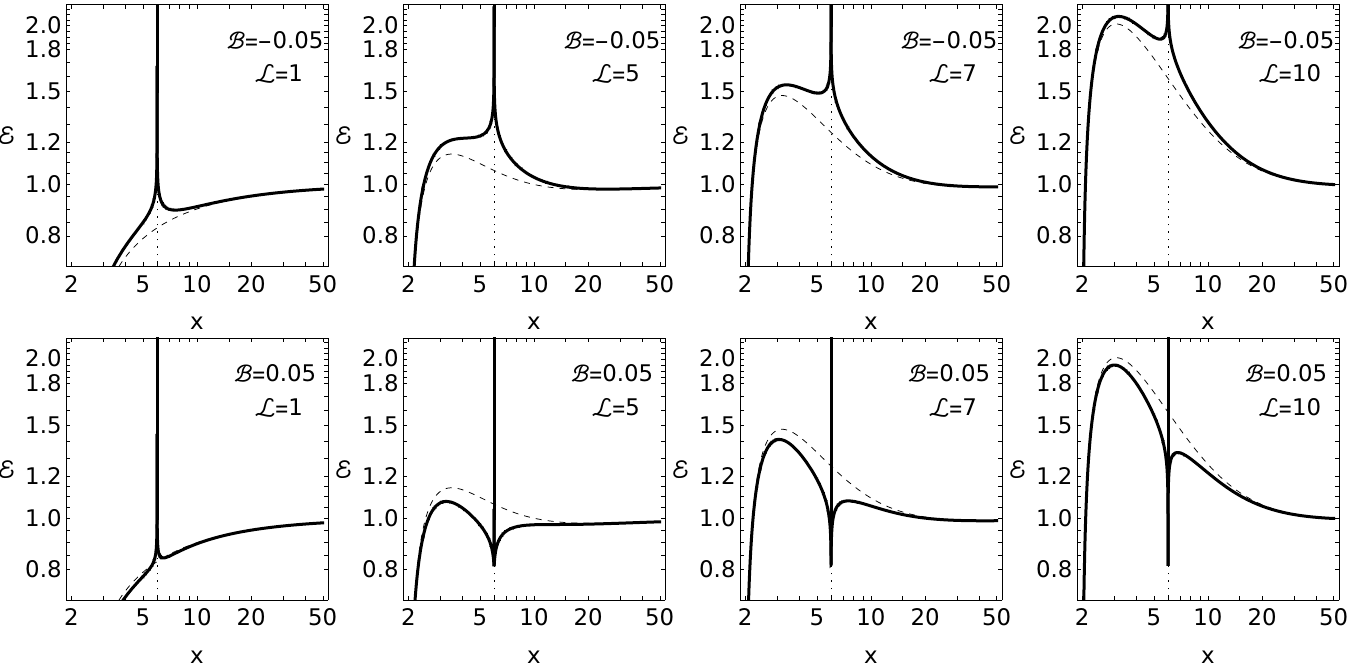}
\caption{Effective potential for charged particle dynamic $V_{\rm eff}(x,z=0)$ plotted in equatorial plane (solid thick curved) as function of coordinate $x$ for various values of magnetic parameter $\cb$ and angular momenta $\cl$. The vertical dotted line at $r_0=6$ is marking position of the current loop, where $V_{\rm eff}(x,z=0)$ diverges to positive infinity. The dashed lines are effective potential $V_{\rm eff}(x,z=\infty)$ cross sections at vertical infinity allowing us to see the full $V_{\rm eff}(x,z)$ structure. \label{fig:Veff_Sec}
}
\end{figure*}

\subsection{Effective potential} \label{sec:eff}

Due to the symmetries of the Schwarzschild spacetime (\ref{SCHmetric}) and the current loop magnetic field (\ref{eq:Aphi}), one can easily find the conserved quantities:  the energy and the axial angular momentum of the particle which can be expressed as
\bea
 E &=& - \cp_t = m f(r) \frac{\d t}{\d \tau}, \\
 L &=& \cp_\phi = m \, r^2 \sin^2\theta \, \frac{\d \phi}{\d \tau} + q A_\phi(r,\theta). \label{angmom}
\eea

Let us, for convenience, introduce parameters, energy $\ce$, axial angular momentum $\cl$, and magnetic interaction parameter $\cb$, by the relations
\beq
\ce = \frac{E}{m}, \quad \cl = \frac{L}{m}, \quad \cb = \frac{q I}{m}, \label{ELB}
\eeq
where $I$ is the electric current through the current loop. The magnetic parameter $\cb$ contains contributions from both the current $I$ and the particle-specific charge $q/m$ since in the particle equation of motion, the product $\cb A_\phi$ always appears, and we provided electromagnetic four-potential generated by a unit current in (\ref{eq:Aphi}).

Now one can rewrite the Hamiltonian (\ref{particleHAM}) in the form
\beq
H = \frac{1}{2} f(r) \p_r^2 + \frac{1}{2r^2} \p_\theta^2  + \frac{1}{2} \frac{m^2}{f(r)} \left[ V_{\rm eff}(r,\theta) - \ce^2 \right], \label{HamHam}
\eeq
where $V_{\rm eff}(r,\theta; \cl,\cb)$ denotes the effective potential given by the relation
\beq
V_{\rm eff} (r,\theta) \equiv f(r) \left[ 1 + \frac{(\cl - \cb A_\phi)^2 }{r^2} \right]. \label{VeffCharged}
\eeq
The terms in the parentheses correspond to the central force potential given by the specific angular momentum $\cl$ and electromagnetic potential energy given by the electromagnetic four-potential. The Hamiltonian (\ref{HamHam}) can be divided into dynamical part $H_{\rm D}$ (first two terms containing dynamical momenta $p_r,p_\theta$) and potential part $H_{\rm P}$ (last term).

The effective potential (\ref{VeffCharged}) shows clear symmetry $(\cl,\cb)\leftrightarrow(-\cl,-\cb)$ and since the particle angular momentum $\cl$ and the magnetic interaction parameter $\cb$ are constant during the particle motion, we are allowed to distinguish two separate configurations $\cl>0$ of the particle dynamic:
\begin{itemize}
\item[$-$] {\it Minus repulsive configuration $\cb<0$} where the Lorentz force is repulsive away from the current loop and we have a peak in the $V_{\rm eff} (r,\theta)$ function, see Fig.~\ref{fig:Veff} (left). The Lorentz force is repulsive acting against BH gravitation pull above the current loop $r>r_0$, while the Lorentz force is attractive towards BH below the current loop $r<r_0$. This configuration is equivalent to $\cl<0, \cb>0$.
\item[+] {\it Plus attractive configuration $\cb>0$} where the Lorentz force is attractive to the current loop and we have a pit in the $V_{\rm eff} (r,\theta)$ function close to the current loop; see Fig.~\ref{fig:Veff} (right). Although we have a depression close to the current loop, we will have a peak exactly at $r=r_0,\theta=\pi/2$ due to the fact that the electromagnetic four-potential $A_\phi(r,\theta)$ (\ref{eq:Aphi}) goes to positive infinity at the current loop. The Lorentz force is attractive because it acts with the gravitational pull of BH above the current loop $r>r_0$, while the Lorentz force is repulsive from BH below the current loop $r<r_0$. This configuration is equivalent to $\cl<0, \cb<0$.
\end{itemize}
In this article, without loss of generality, we use the positive angular momentum of a particle $\cl>0$, while the magnetic parameter $\cb$ can be both positive or negative. The attractive configuration $\cb>0$ is of particular astrophysical interest, since due to the attractive Lorentz force nature, the charged particles can accumulate close to the current loop and hence generate a cumulative ring current. There is an open question whether the presence of charged particle cumulative ring current could allow the existence of a self-generating toroidal plasma pinch structure. 

The effective potential $V_{\rm eff}$ (\ref{VeffCharged}) shares the background symmetries of the spacetime metric and those of the superposition magnetic field. It is independent of the coordinate $\phi$ and diverges at the event horizon $r_{\rm p} = 2 M$ and at the current loop $r=r_0,\theta=\pi/2$. We exclude the regions of the horizon and divergent points from our considerations; the effective potential is positive everywhere outside the horizon. Due to reflective background symmetries, we can focus on the region $\theta\in(0,\pi/2)$ only. In Fig.~\ref{fig:Veff_Sec} we plot the effective potential as a function of $x$ for the positive and negative values of the magnetic field parameter $\cb$ using standard Cartesian coordinates $x$-$z$
\beq 
x = r \, \cos{\phi} \, \sin{\theta}, \,\,\, y = r \,\sin{\phi} \, \sin{\theta}, \,\,\, z = r \, \cos{\theta}.
\eeq
The electromagnetic four potential $A_\phi$ (\ref{eq:Aphi}) is divergent at the current loop position $r_0$. The effective potential $V_{\rm eff}$ (\ref{VeffCharged}) has divergence $+\infty$ at the current loop position $r_0$ as well due to the presence of square for any parameter $\cb$ or angular momentum $\cl$. It is not possible to have a stable charged particle circular orbit exactly at the current loop. However, as we can see in Fig.~\ref{fig:Veff_Sec} for positive magnetic parameter $\cb>0$ we can have a local minimum in the neighborhood of the current loop for some values of angular momenta $\cl$. Here, the square term in $V_{\rm eff}$ (\ref{VeffCharged}) will be minimized and the charged particle can be effectively trapped in orbit around the current loop, yet the exact position of the current loop $r_0$ is impossible to reach.

\begin{figure*}
\centering
\includegraphics[width=\textwidth]{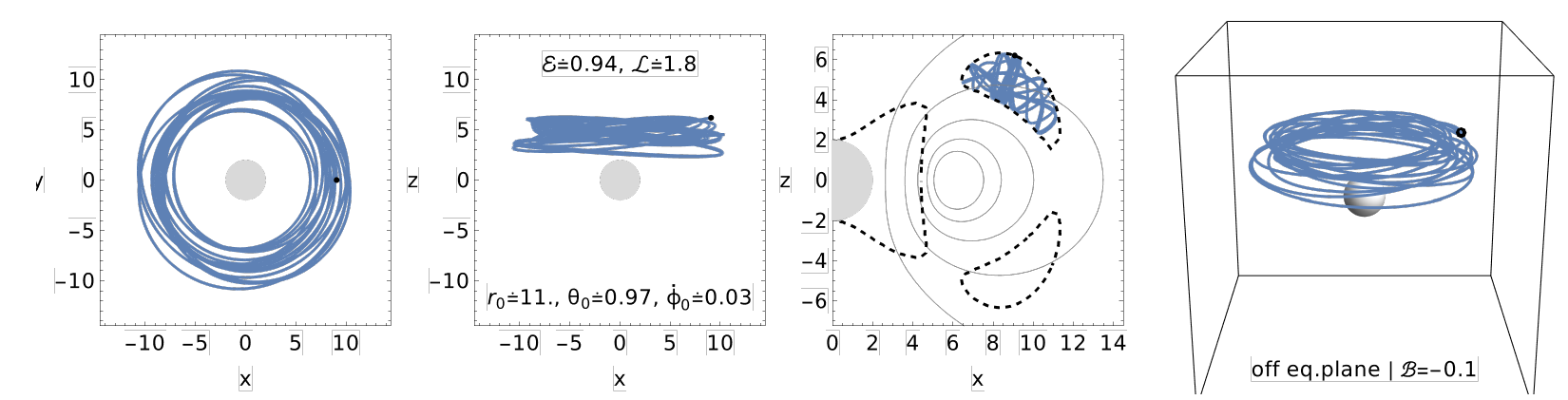}
\includegraphics[width=\textwidth]{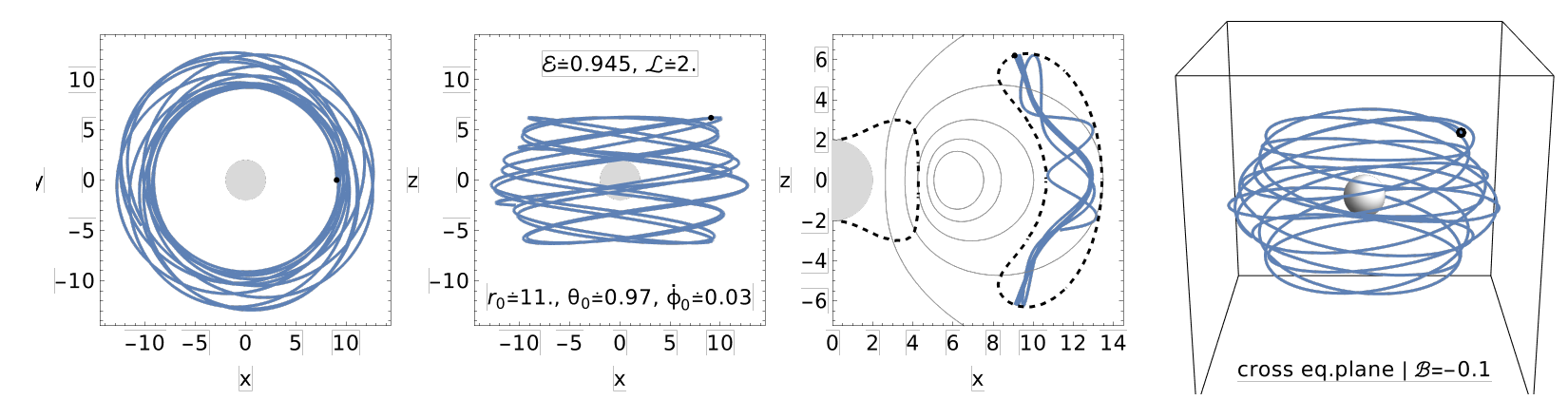}
\caption{
Charged particle motion in BH magnetosphere above the current loop $r>r_0$ where the magnetic field is close to dipole field shape. Two bounded trajectories forming structures similar to radiation belts are plotted: particles with particle off equatorial plane confinement (upper row) and particles crossing equatorial plane (lower row). We plotted each trajectory using various different views (from left to right): from top; from side; the axial projection where $\phi$ coordinate is neglected; and full 3D view. 
The gray disk represents BH, solid red curve is particle trajectory, black dot initial particle position, thick black dashed curves are forming particle energetic boundary, gray curves are lines of magnetic field.
\label{fig:tra:out}
}
\end{figure*}

\begin{figure*}
\centering
\includegraphics[width=\textwidth]{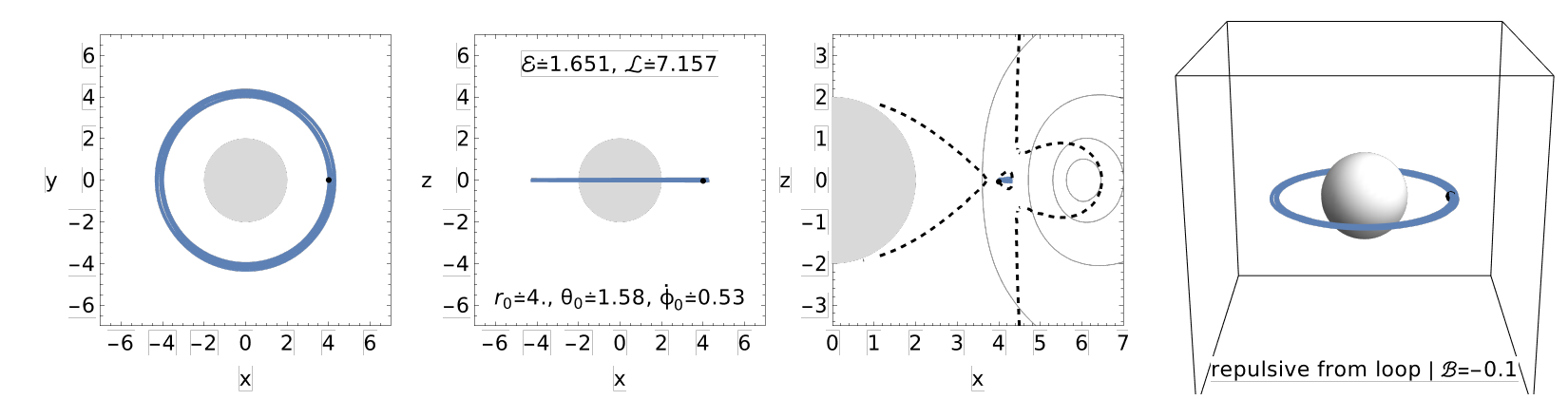}
\includegraphics[width=\textwidth]{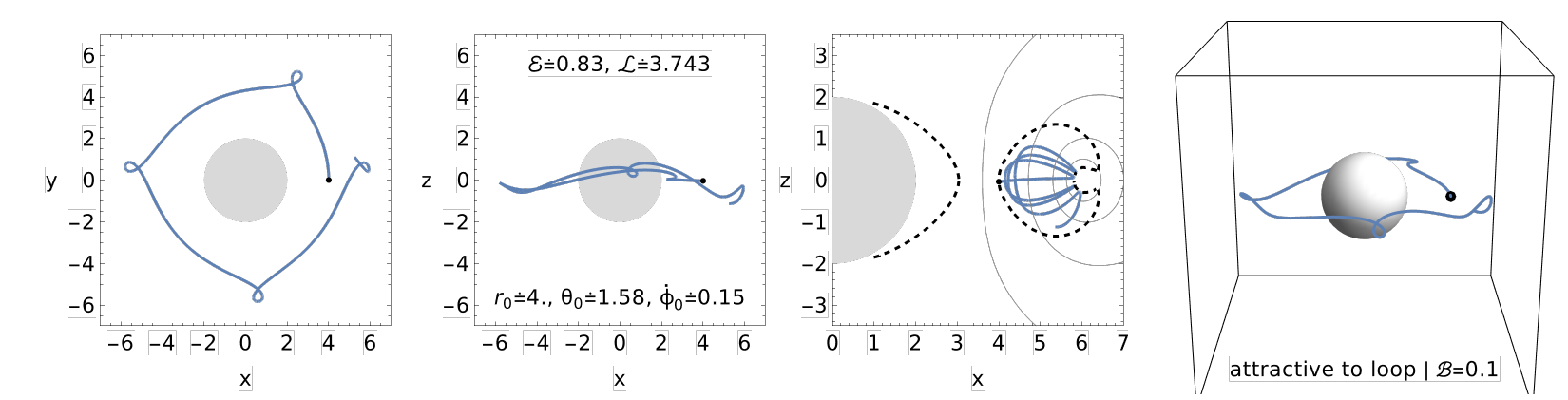}
\caption{Charged particle dynamic in BH magnetosphere below the current loop $r<r_0$ where the magnetic field is close to the uniform field shape at last in restricted region close to equatorial plane. The Lorentz force is attractive on upper row of figures and particle is orbiting BH faster then on circular geodesic; the  Lorentz force is repulsive on lower row of figures when the trajectory curls are formed. 
\label{fig:tra:in}}
\end{figure*}

\begin{figure*}
\centering
\includegraphics[width=\textwidth]{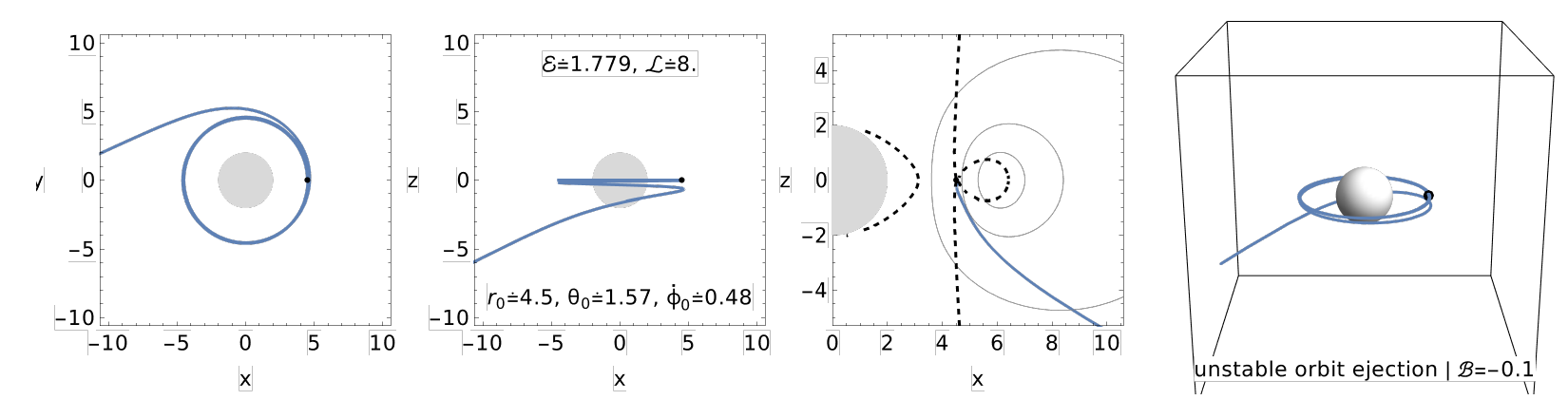}
\includegraphics[width=\textwidth]{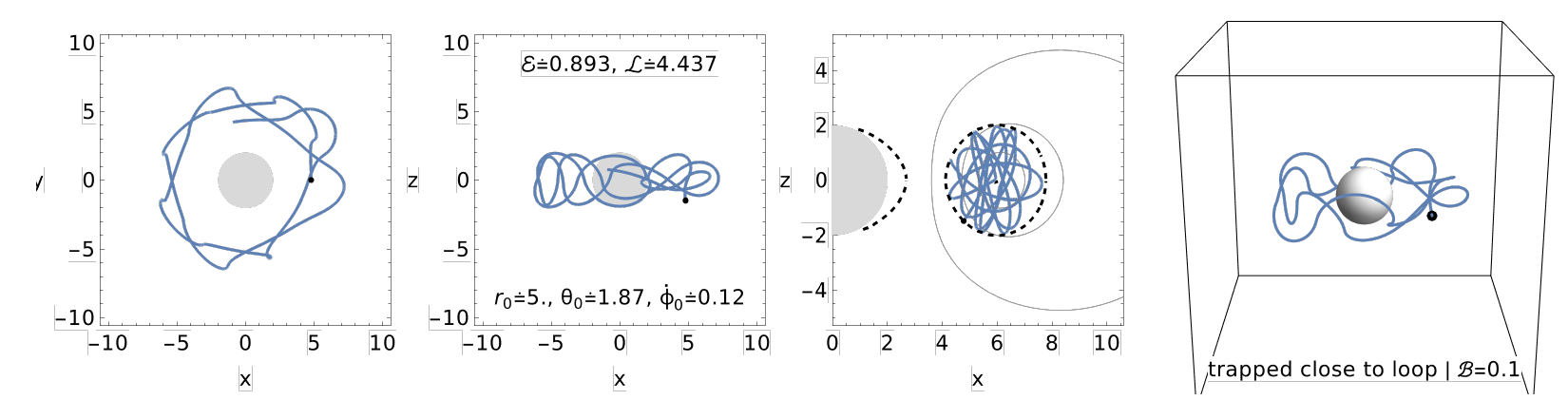}
\includegraphics[width=\textwidth]{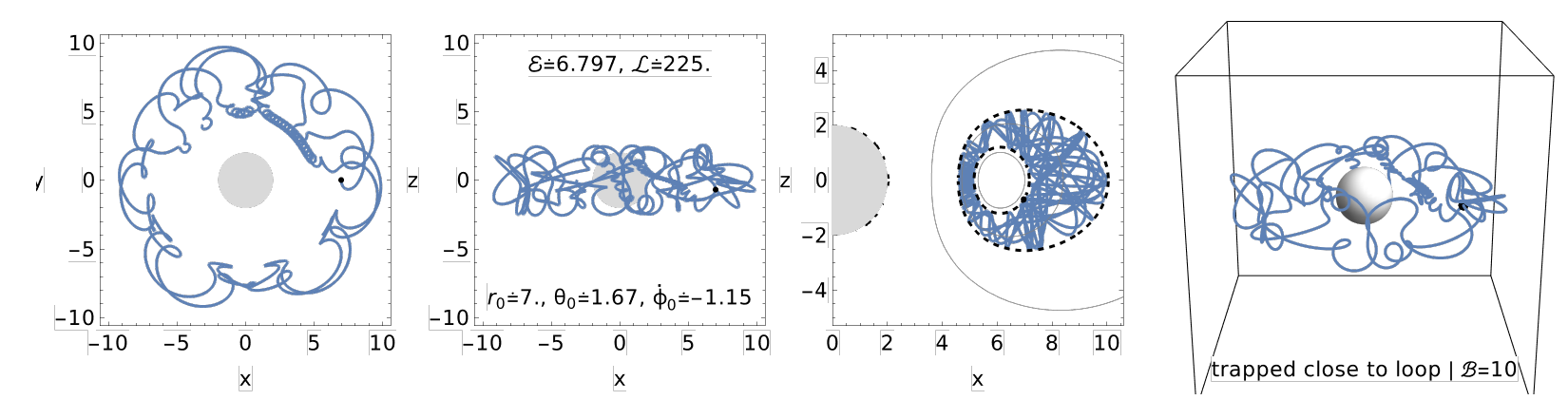}
\caption{Charged particle dynamic in BH magnetosphere close to the current loop $r\sim{}r_0$. Charged particle is ejected from BH magnetosphere on upper row of figures; charged particle is trapped is region close to the current loop on lower row of figures. Figure description similar to the Fig.~\ref{fig:tra:out}.
\label{fig:tra:close}}
\end{figure*}

\subsection{Charged particle trajectories} 

Charged particle trajectories can be obtained by numerical integration of equations of motion (\ref{Ham_eq}) and we have plotted representative trajectories in Figs.~\ref{fig:tra:out}-\ref{fig:tra:close}. Outside of the current loop, for radii $r>r_0$, we observed a charged particle dynamic similar to that of the dipole magnetic field; see Fig.~\ref{fig:tra:out}. The motion of charged particles in the dipole magnetic field around the compact object has already been explored in \cite{Vrb-Kol-Stu:2024:EPJC:}. In Fig.~\ref{fig:tra:out} we demonstrate the existence of a trapped orbit close to the minima of effective potential $V_{\rm eff}(r,\theta)$ in the equatorial plane. We also demonstrate the formation of off-equatorial minima in the case when the minimum of $V_{\rm eff}(r,\theta)$ in the equatorial plane becomes a saddle point. For larger radii above, the current loop $r\gg{}r_0$, or if the current loop radius is relay small, the charged particle will move in an almost perfect dipole magnetic field \cite{Vrb-Kol-Stu:2024:EPJC:}. For the test particle dynamic within the current loop radius $r<r_0$ we have dynamics in uniform magnetic field configuration, finally in the vicinity of the equatorial plane, see Fig.~\ref{fig:tra:in}. If the current loop radius will be extended to $r_0\gg{}2M$ or even moved to infinity, then a uniform magnetic field will be generated everywhere around BH. The dynamic of the charged test particle in a uniform magnetic field has already been explored in great detail in \cite{Kol-Stu-Tur:2015:CLAQG:} and in all the following articles. In Fig.~\ref{fig:tra:in} we plotted the trajectory for the attractive Lorentz force when the charged particle orbits the BH faster than in the circular geodesic and the repulsive Lorentz force when trajectory curls are formed due to backward movement in the orbital dynamic. 

A new and quite interesting dynamic can be obtained when the charged particle is moving close and around the current loop $r\sim{}r_0$, see Fig.~\ref{fig:tra:close}. Here we can see not only charged particle being expelled from loop, but also another important case of dynamic when the charged particle is trapped on helical trajectory around the current loop. This trapped trajectory can be easily understood since, for positive configuration $\cb>0$, the Lorentz force attracts charged particles toward the current loop. We have a depression formed around the current loop in the effective potential function $V_{\rm}(r,\theta)$ where the charged particle can be trapped; see Fig.~\ref{fig:Veff}. The current loop itself is not accessible by the charged particle dynamic as there is always infinite a peak due to the four-potential $A_\phi$ divergence.

\begin{figure*}
\centering
\includegraphics[width=\textwidth]{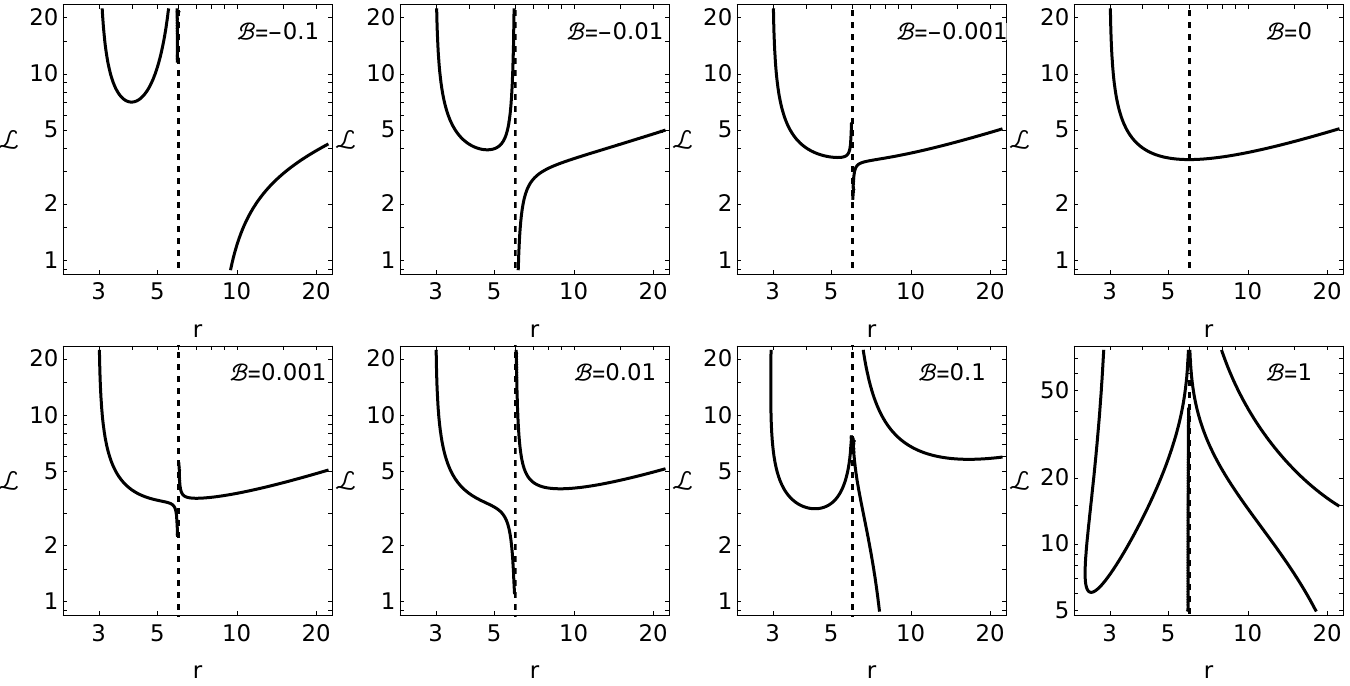}
\caption{
Angular momentum for charged particle circular orbit as a function of radius $\cl_{\rm C}(r)$ in the equatorial plane is plotted as thick black curve. Positions of stationary points of the effective potential $V_{\rm eff}(r,\pi/2)$ are places where the circular orbit can be located; $V_{\rm eff}$ minimum is giving stable circular orbit, while maximum and saddle point unstable one. For given angular momentum $\cl$, which is constant of particle motion and will be imagined as horizontal line, circular orbits with different radii can be obtained as cross sections with $\cl_{\rm C}(r)$ curve. 
The stable circular orbits exists when function $\cl_{\rm C}(r)$ is increasing with radius ($\d\cl_{\rm C}/\d{}r>0$) and unstable circular orbits are when function $\cl_{\rm C}(r)$ is decreasing, giving ISCO position $\d\cl_{\rm C}/\d{}r=0$, see \cite{Abr-Fra:LRR:2013:}.
Vertical dotted line represent the position of current loop, where $\cl_{\rm C}(r)$ diverge for $\cb\neq0$. For magnetic parameter $\cb=0.1$ we can also see there can exist two different unstable orbits at the same position but with different angular momentum and orbital speed. 
\label{fig:Lcirc}
}
\end{figure*}

\subsection{Stability of circular orbits} 

Radiation belts that form the magnetosphere around central objects are created by trapped charged particles that collectively occupy distinct regions of the phase space corresponding to specific energy levels given by the $V_{\rm eff}(r,\theta)$ boundary. We would like to generate radiation belts in the BH magnetosphere generated by current loop and even to explore interaction between the current loop electromagnetic field and charged particles that do exist on helical trajectories presented in Fig.~\ref{fig:tra:close}. The bounded orbits with the lowest energies are stable circular orbits which are always located in the center of the radiation belt.

The effective potential represents an energetic boundary for the charged particle dynamic
\beq
 \ce^2 = V_{\rm eff} (r,\theta). \label{MotLim}
\eeq
The stationary points of $V_{\rm eff}(r,\theta)$, which define the minima (maxima, saddle points), are given by the derivatives 
\beq
  \partial_r V_{\rm eff}(r,\theta) = 0, \quad \partial_\theta V_{\rm eff}(r,\theta) = 0.  \label{extrem}
\eeq
which are related to particle stable (unstable) circular orbit.

From condition (\ref{extrem}) the angular momentum for a charged particle on circular equatorial orbit around spherically symmetric BH in general $A_\phi$ four-potential can be expressed
\bea
 \cl_{\rm C}(r) =  \frac{1}{{r f'-2 f}} && \bigg( 
 r A_{\phi} f'+ f r A_{\phi}' -2 f A_{\phi} \nonumber \\
 && \pm r \sqrt{2 f r f'-r^2{f'}^2+f^2 {A_{\phi}'}^2} 
 \bigg) \label{Lcirc}
\eea
where the prime $'$ denotes derivative with respect to the coordinate $r$ for $\theta=\pi/2$. Extrema of $\cl_{\rm C}(r)$ will tell us more about the circular orbit radial stability for which $\partial_r{\cl_{\rm C}}>0$ and innermost stable circular orbit (ISCO) are located at $\partial_r{\cl_{\rm C}}=0$.  The circular orbit angular momentum function $\cl_{\rm C}(r)$ is plotted in Fig.~\ref{fig:Lcirc}. 

The Fig.~\ref{fig:Veff_Sec} and Fig.~\ref{fig:Lcirc} are in close relation to each other, where the Fig.~\ref{fig:Lcirc} for circular orbit angular momenta provides position of effective potential extrama from Fig.~\ref{fig:Veff_Sec}. One should keep in mind, that $V_{\rm eff}$ diverges to $+\infty$ at the current loop $r_0$. 
In the previously explored charged particle dynamic around Schwarzschild BH with the uniform magnetic field \cite{Kol-Stu-Tur:2015:CLAQG:}, the effective potential $V_{\rm eff}$ has minima in the equatorial plane only. For the case of magnetic field generated by current loop discussed in this work, the situation is different, and there are also effective potential minima located above (below) the equatorial plane. Such off-equatorial minima has been already reported in dipole magnetic field case, see \cite{Vrb-Kol-Stu:2024:EPJC:} and this result is not surprising since the current loop magnetic field can be approximated by uniform below the current loop $r<r_0$ while it can be approximated by dipole field above the current loop $r>r_0$.

\begin{figure*}
\centering
\includegraphics[width=\textwidth]{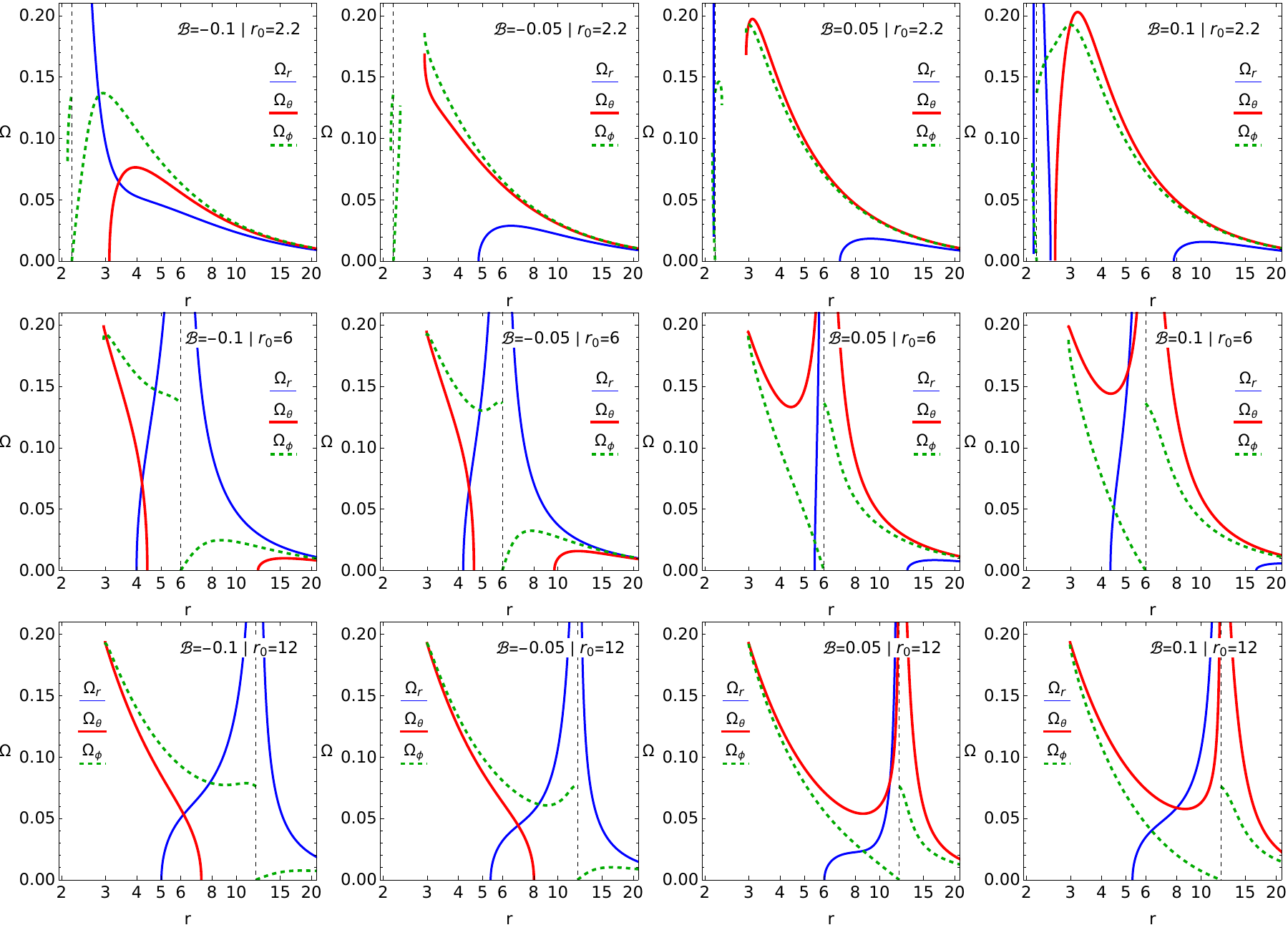}
\caption{Fundamental frequencies of charged particle oscillation on a circular orbit in the equatorial plane. We chose three representative loop positions $r_0\in\{4,6,12\}$, each with a different magnetic field with $k$ interaction parameter so that the various frequency behavior can be covered. Such frequencies are useful for determining charged particle orbital stability while showing different orbital dynamics time scales.
There is a close connection to Fig.~\ref{fig:ISCO}, where stability condition $\Omega_r^2>0$ is presented.
\label{fig:Freq}
}
\end{figure*}

\subsection{Particle fundamental frequencies} 

If a test particle is slightly displaced from the equilibrium position at a stable circular orbit, the particle starts to oscillate around the equilibrium position, realizing epicyclic motion governed by linear harmonic oscillations. 

We can separate the charged particle Hamiltonian (\ref{HamHam}) into dynamical (with momenta $p_r,p_\theta$) and potential part $H_{\rm p}$ with coordinate functions only and then the radial and vertical angular frequencies in the equatorial plane are given by
\bea
 \omega^2_{\mir}  &=&  \frac{1}{2 g_{rr}} \left( \frac{\partial^2  H_{\rm p}}{\partial r^2} \right), \label{omegaR} \\
 \omega^2_{\mit}  &=&  \frac{1}{2 g_{\theta \theta}} \left( \frac{\partial^2  H_{\rm p}}{\partial \theta^2}  \right). \label{omegaT}
\eea
The more general formula for off-equatorial perturbations can be found in \cite{Kol-Sha-Tur:2023:EPJC:}. In equatorial plane the frequencies as measured by local observer can be expressed for four-potential (\ref{eq:Aphi}) as
\def\tAp{\cb A_\phi}
\bea
\omega_{r}^2 = && \frac{f}{r^4} \bigg[
3 (\cl-\tAp)^2 +r(\cl-\tAp)(4 \tAp'-r \tAp'')  \nonumber\\ 
&&  +r^2{\tAp'}^2 + \frac{r^2 (f f''-2 {f'}^2) \left(\left(\cl-\tAp\right)^2+r^2\right)}{2 f^2}
\bigg], \\
\omega_{\theta}^2 = && \frac{1}{r^4} \Big[ (\cl - \tAp)^2 -(\cl-\tAp) \tAp^{**}  + {\tAp^{*}}^2 \Big],
\eea
where $\cl=\cl(r)$ is circular orbit angular momentum (\ref{Lcirc}). We have used $f'=\partial_r f(r)$ and $f''=\partial^2_{rr}f(r)$ for lapse function (\ref{fun_f}) and $A_\phi'={\partial_r}A_\phi$, $A_\phi''=\partial^2_{rr}A_\phi$, $A_\phi^{*}={\partial_\theta}A_\phi$, $A_\phi^{**}=\partial^2_{\theta\theta}A_\phi$ evaluated in equatorial plane $\theta=\pi/2$ (see Eqs.~(\ref{eq:dAphi}\,--\,\ref{eq:dArr}) for the exact expressions).

There exists the third fundamental angular frequency of the epicyclic particle motion, namely the Keplerian (axial) frequency $\omega_\mip$, given by
\beq
 \omega_{\mip} \equiv u^\phi = g^{\phi\phi} \left( \cl - q A_\phi \right). \label{omegaP}
\eeq
In contrast to the neutral case, there exists another angular frequency -- so-called Larmor frequency -- for charged particles, which is associated with the magnetic field only, and given by a relation
\beq
\omega_{\rm L} = \frac{q}{m} |\textbf{B}|,
\eeq
where $|\textbf{B}|$ can be found by Eq. (\ref{magnitude}). 

The characteristics and ratios of the fundamental frequencies $\omega_\mir, \omega_\mit, \omega_\mip$, are useful to differentiate various shapes of the epicyclic orbits of charged particles and their stability. In the classical Newtonian theory of gravitation, all frequencies are equal, $\omega_\mir=\omega_\mit=\omega_\mip$, giving the ellipse as the only possible bounded trajectory of a test particle around a gravitating spherically symmetric body. For uncharged particles moving around a Schwarzschild BH, the relation $\omega_\mir<\omega_\mit=\omega_\mip$ holds, and there exists a periapsis shift for bounded elliptic-like trajectories implying the effect of relativistic precession that increases with decreasing radius of the orbit as the strong gravity region is entered \cite{Ste-Vie:1999:PHYSRL:}. 

Frequencies for charged particles orbiting magnetized BH one gets in general $\omega_\mir\neq\omega_\mit\neq\omega_\mip$, see \cite{Kol-Stu-Tur:2015:CLAQG:,Tur-Stu-Kol:2016:PRD:,Kol-Tur-Stu:2017:EPJC:}. In the Earth magnetosphere for realistic astrophysically relevant values one will get $\omega_{\rm L} \gg \omega_\theta \gg \omega_\phi$, where $\omega_{\rm L}$ represents the gyration motion along the magnetic field line, the $\omega_{\theta}$ bounce motion between the magnetic poles and the $\omega_{\omega}$ particle drift around the central object. For ions in Earth magnetosphere, all there frequencies have completely different magnitudes with the difference between the individual frequencies up to 4 orders of magnitude \cite{Kos-Kil:2022:BOOK:}. In this work, we use the test-particle fundamental frequencies to analyze orbital stability. An orbit is stable when $\omega^2 > 0$ and the frequency $\omega$ is real. Each frequency, $\omega_r, \omega_\theta, \omega_\phi$, corresponds to stability in the $r, \theta, \phi$ directions, respectively. The motion of charged particles is periodic when $\omega^2 > 0$ and $\omega$ has a real solution. For $\omega^2 < 0$, the fundamental frequency $\omega$ becomes complex, leading to a runaway solution (escape or collapse into the black hole).

The locally measured angular frequencies $\omega_\mir$, $\omega_\mit$, $\omega_\mip$ and $\omega_{\rm L}$ are connected to the angular frequencies measured by the static distant observers, $\Omega_{\beta}$, by the gravitational redshift transformation
\beq
 \Omega_{\beta} = \frac{\d \aaa_\beta}{\d t} = \omega_{\beta} \, \frac{\d \tau}{\d t} = \omega_{\beta} \, \frac{f(r)}{ \ce(r)},
\eeq
where $(f(r) / \ce(r) ) $ is the redshift coefficient, given by the  function $f(r)$ and the particle specific energy at the circular orbit $\ce(r)$. In Fig.~\ref{fig:Freq} we plot the frequencies $\Omega(r)$ for charged particles on perturbed circular orbits of radius $r$ in the equatorial plane, with current loops located at three representative positions. Although the behavior of $\Omega(r)$ is complex, several key features can be identified:
\begin{itemize}
\item The radial frequency $\Omega_r$ increases near the loop with increasing magnetic field magnitude. A large region of radial instability, $\Omega_r^2 < 0$, occurs above the current loop in the Lorentz-repulsive case ($\cb > 0$), due to the presence of a peak at $r_0$ in the effective potential $V_{\rm eff}$.
\item A vertical instability, $\Omega_\theta^2 < 0$, appears near the loop in the Lorentz-repulsive case ($\cb < 0$). In contrast, for Lorentz attraction towards the loop, the vertical frequency $\Omega_\theta$ increases near the loop.
\item A steep change in $\Omega_\phi$ occurs at the radius $r_0$, corresponding to a reversal of the Lorentz-force orientation. The Lorentz force points in opposite directions above and below the loop, either towards or away from the black hole. Depending on orientation, the Lorentz force adds to or subtracts from the black hole’s gravitational pull, which must be balanced by a larger (smaller) centrifugal force generated by faster (slower) orbital motion in the $\phi$ direction.
\end{itemize}

\begin{figure*}
\centering
\includegraphics[width=\textwidth]{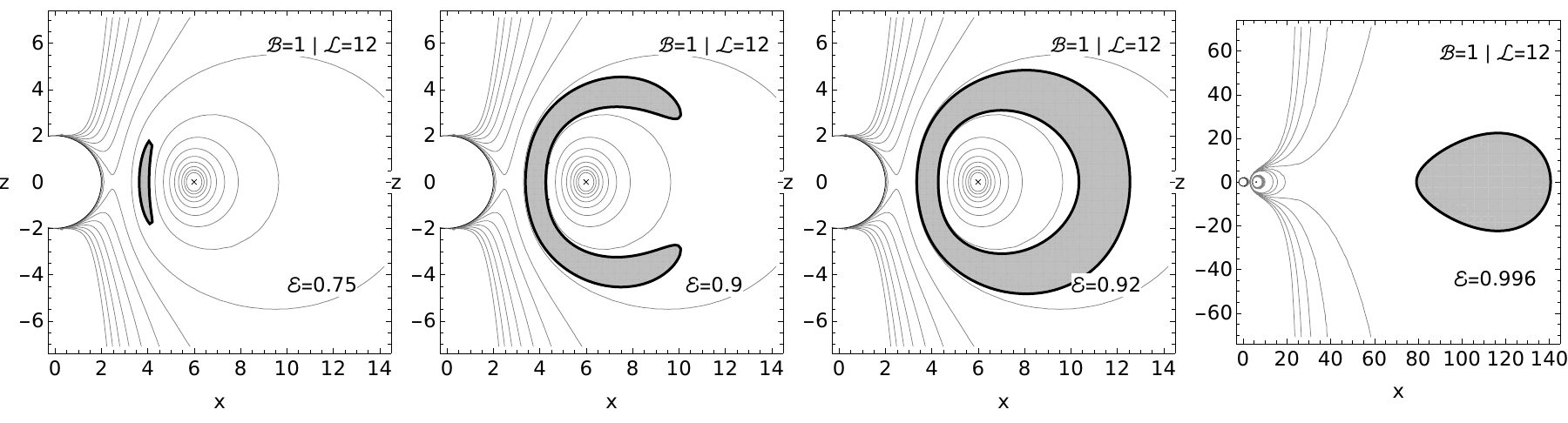}
\includegraphics[width=\textwidth]{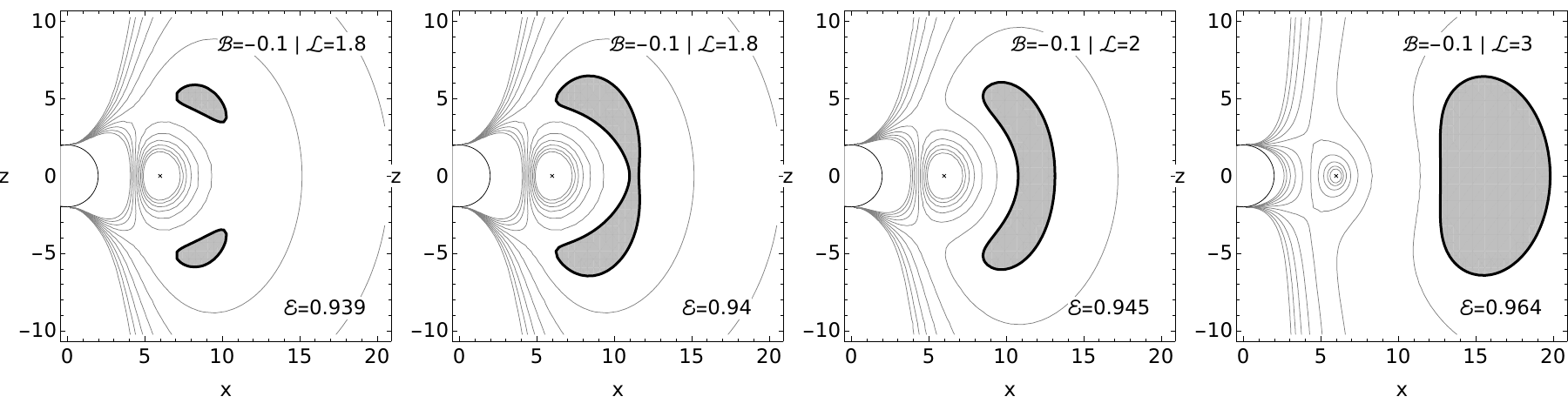}
\caption{Positions and sizes of allowed regions (radiation belts) for charged particle dynamic in BH magnetosphere generated by current loop. The upper row of the figures represents situation when the Lorentz force is attractive towards to the current loop ($\cb=1$) and the allowed regions are forming around the current loop. The lower row of figures is for Lorentz force repulsive from the current loop ($\cb=-0.1$) and where we can have of-equatorial allowed regions.
Standard regions in equatorial plane can be formed in both $\cb>0$ and $\cb<0$ cases.
\label{fig:belts}
}
\end{figure*}

\begin{figure*}
\centering
\includegraphics[width=\textwidth]{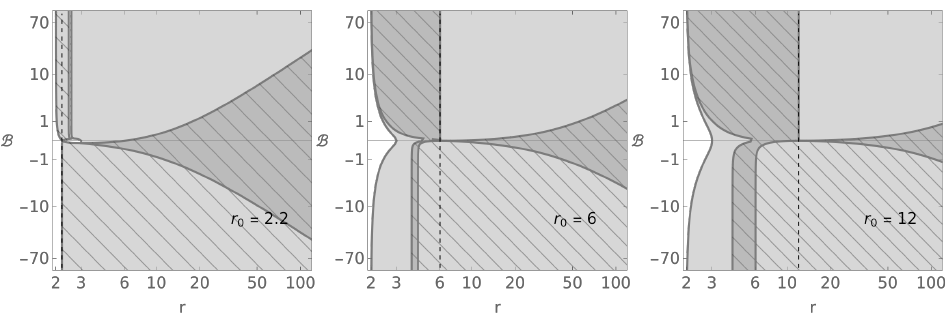}
\includegraphics[width=\textwidth]{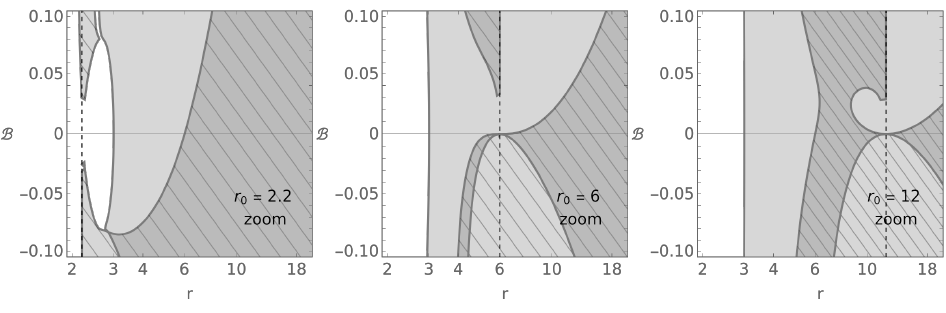}
\caption{Region of radially stable circular orbits, $\Omega_r(r,\cb) > 0$ (hatched gray), and region of vertical stability, $\Omega_\theta(r,\cb) > 0$ (non hatched gray). In the overlapping region, the minima of the effective potential $V_{\rm eff}(r,\theta)$ can exist, and stable circular orbits occur there (darker hatched gray). No stable orbits are allowed in the white region below the Innermost Stable Circular Orbit (ISCO) (dark gray curves) and outside the region allowing the existence of charged-particle stable circular orbits (gray area). The position of the generating current loop, $r_0$, is plotted as a dashed vertical line. The radial ISCO position is also given by the extrema of $\cl_{\rm C}(r)$ plotted in the previous Fig.~\ref{fig:Lcirc}. 
\label{fig:ISCO}
}
\bigskip
\includegraphics[width=\textwidth]{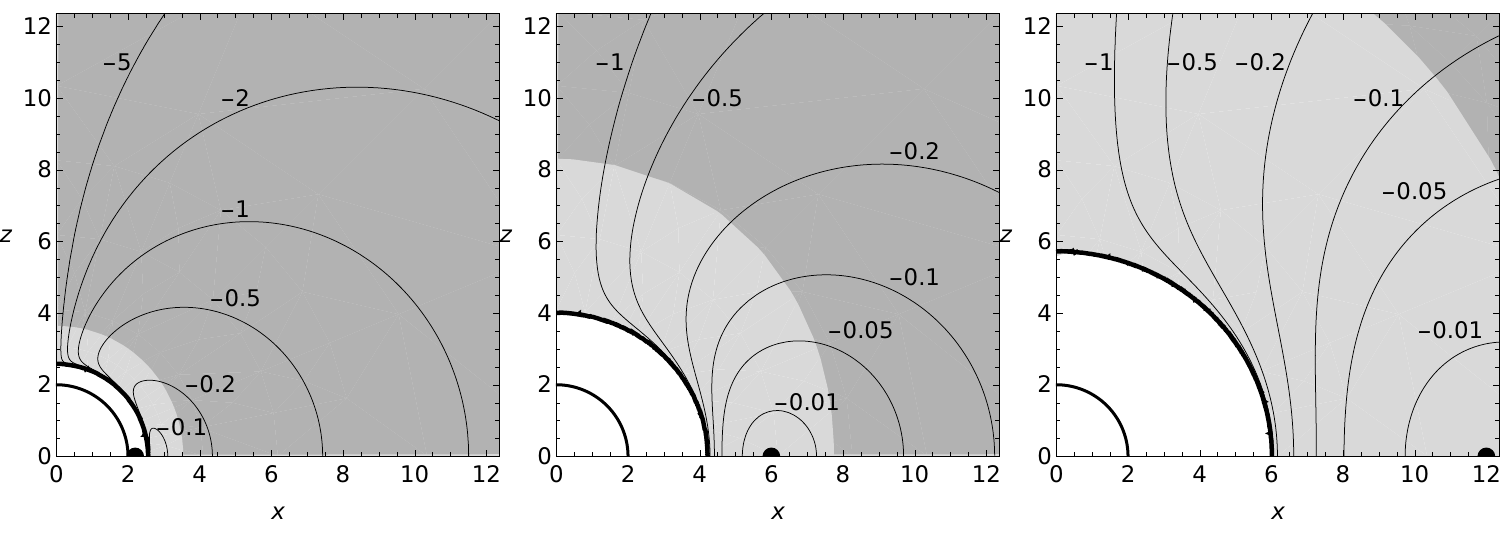}
\caption{Regions admitting the existence of off–equatorial circular orbits exist only for repulsive Lorentz force $\cb<0, \cl>0$ (equivalent to $\cb>0, \cl<0$ ). In the white regions above the black hole horizon, $r > r_{\rm p}$, no stationary points of the effective potential $V_{\rm eff}$ are present. In the light–gray regions, $V_{\rm eff}$ may exhibit saddle points for specific values of the angular momentum $\cl$, corresponding to the presence of unstable circular orbits. In the dark–gray regions, local minima of $V_{\rm eff}$ occur, allowing for stable circular orbits. For each value of the magnetic parameter $\cb$, the stationary points of $V_{\rm eff}$ are indicated by black curves, with the corresponding magnitude of $\cb$ specified by the adjacent labels.
\label{fig:offeqmin}
}
\end{figure*}

\subsection{Existence of radiation belts}

Radiation belts are regions populated by trapped charged particles, typically electrons and ions, confined by the combined influence of the gravitational field of a central object and its magnetic field. Within these belts, the particles undergo rapid gyromotion along circular Larmor orbits perpendicular to the magnetic field lines. Simultaneously, they bounce along the magnetic field lines in the vertical (poloidal) $\theta$-direction and slowly drift azimuthally in the orbital (toroidal) $\phi$-direction around the central object. From the shape of the effective potential $V_{\rm eff}(r,\theta)$, it is evident that the bound orbit with the lowest energy corresponds to a stable circular trajectory located in the equatorial plane, at the minimum of $V_{\rm eff}$. In astrophysically relevant scenarios, the situation is considerably more complex. By analogy with the Earth and planetary magnetospheres, electromagnetic waves propagating through the magnetosphere are expected to play a significant role in the formation, structure, and location of radiation belts. 

The stability of circular orbits is crucial for the existence of radiation belts—regions surrounding a gravitating object. Each such structure is centered around a circular orbit, as it represents the configuration of lowest energy. All other bound orbits of charged particles must possess energies exceeding that of the corresponding circular orbit. In regions where no circular orbit exists for charged particles, such as below the ISCO, a radiation belt structure cannot form. This analysis applies to collisionless plasmas, which can be modeled as ensembles of non-interacting charged particles. Once particle interactions are introduced, the dynamics become significantly more complex as a result of the emergence of collective effects, such as pressure, leading to thick toroidal structures best described by GRMHD. The upper limit on the energy of trapped particles corresponds to the escape condition, i.e., the energy at which particles can reach infinity—trapped particle energy must be lower than the minimal value of $V_{\rm eff}$ at infinity.

An example of allowed regions (radiation belts) for the dynamic of charged particles in the BH magnetosphere generated by a current loop is plotted in Fig.~\ref{fig:belts}. 
The upper row of the figures represents the situation when the Lorentz force is attractive towards the current loop ($\cb=1$) and there is a dip in the effective potential $V_{\rm eff}(r,\theta)$ function close to the loop, but due to the divergence of the four-potential electromagnetic function $A_\phi(r,\theta)$ at $r=r_0,\,\theta=\pi/2$ there will always be a peak at the current loop position. The angular momentum was kept the same value $\cl=12$, meaning that the shape of the effective potential function $V_{\rm eff}(r,\theta)$ does not change in the four upper figures. There is a close connection to Fig.~\ref{fig:Lcirc} where we plotted the position of the minima of the function $V_{\rm eff}(r,\pi/2)$. We see how the particle-allowable regions are changing with increasing energy in the first three figures, slowly encircling the current loop. 
The lower row of the figures in Fig.~\ref{fig:belts} represents a situation with Lorentz force repulsive from the loop. Here we show the formation of two symetrical off-equatorial minima in effective potential $V_{\rm eff}(r,\theta)$, which can for large energies be connected in a "banana" shape structure. 
Standard regions of allowed particle motion can be formed in both the $\cb>0$ and $\cb<0$ cases in the equatorial plane above the current loop (right column).

As we can see from Fig.~\ref{fig:ISCO}, stable circular orbits for charged particles do not exist for every radius $r$ when we take into account the magnetic parameter $\cb$ and the loop position $r_0$. The existence of stable circular orbits becomes very complex for low values of the magnetic parameter $|\cb|$. For large values, $|\cb| > 1$, a consistent behavior is obtained, regardless of the value of $\cb$. For $\cb > 1$, where gravitational effects become weaker, there is a large stability region extending from the BH horizon to the current loop, while for $\cb < 0.1$ the stability region below the current loop is limited. A stability region also exists above the current loop, but it is located farther away as the magnetic field magnitude $|\cb|$ increases.

The off–equatorial stationary points of the effective potential $V_{\rm eff}(x,z)$ are directly related to the stationary points located in the equatorial plane. For instance, a local minimum of the one–dimensional function $V_{\rm eff}(x,z=0)$ may correspond to a saddle point of the full two–dimensional potential $V_{\rm eff}(x,z)$, which in turn gives rise to the existence of two symmetric minima situated above and below the equatorial plane. In Fig.~\ref{fig:offeqmin} we show the positions of the off–equatorial circular orbits (i.e., stationary points of $V_{\rm eff}(x,z)$) together with their stability properties. No off–equatorial circular orbits are found inside the current loop radius, i.e., for $r<r_0$. This result is consistent with the case of a uniform magnetic field \cite{Kol-Stu-Tur:2015:CLAQG:}, since a uniform field provides a good approximation to the inner region of the current loop solution. The off–equatorial minima of the effective potential can occur only in the Lorentz–repulsive regime, $\cb<0$, and only outside the current loop radius, $r>r_0$. This outcome is again consistent with previous findings obtained for the dipole magnetic field configuration \cite{Vrb-Kol-Stu:2024:EPJC:}, since the dipole field represents an adequate approximation of the magnetic structure in the region $r>r_0$.

\begin{figure*}
\includegraphics[width=0.7\textwidth]{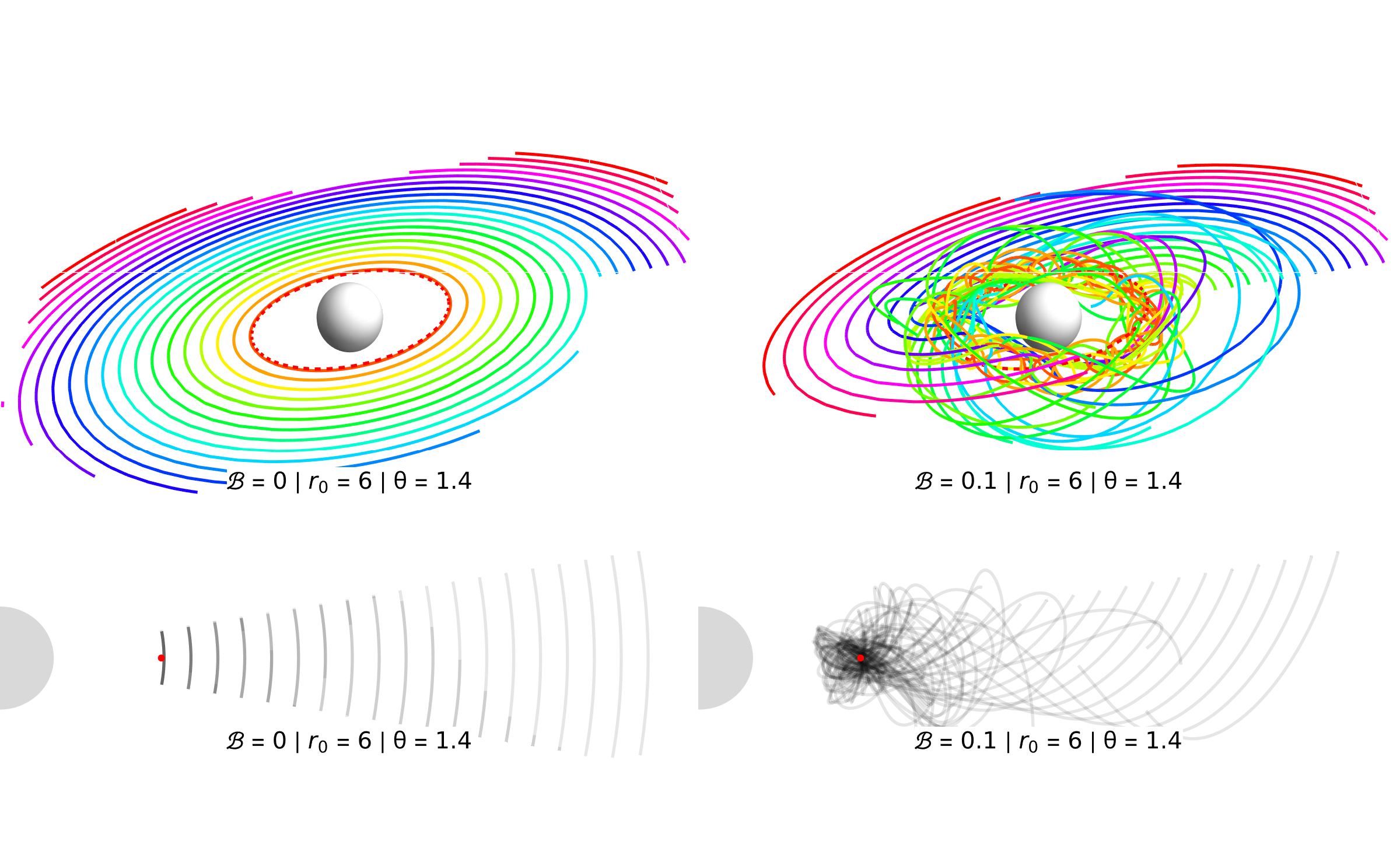}
\caption{
Ionization of thin Keplerian accretion disks around Schwarzschild BH with magnetic current loop in equatorial plane. Before ionization the neutral particles are moving on circular orbit with the small initial inclination $\theta_0=1.4$ (upper left, $\cb=0$). The charged particle after ionization are reacting to the presence of magnetic field and culminating close to the current loop since the Lorentz force is attractive toward the loop ($\cb=0.1$), see upper right sub-figure. Colors indicate initial radial coordinate in thin disk ($r_0$) and the particles are following circular geodesics with $\cl > 0$ in counterclockwise direction. The gray sphere and half disk indicate BH horizon, and red dot the position of current loop.
Both lower sub-figures are showing the same trajectories as the the upper upper two sub-figures, but both lower sub-figures we the particles toroidal motion in $\phi$ direction has been neglected. The lower gray plots can be seen as visual representation of particle density probability in radiation belts structures around BHs generated by current loop. \label{fig:DISK}
}
\end{figure*}

\begin{figure*}
\centering
\includegraphics[width=\textwidth]{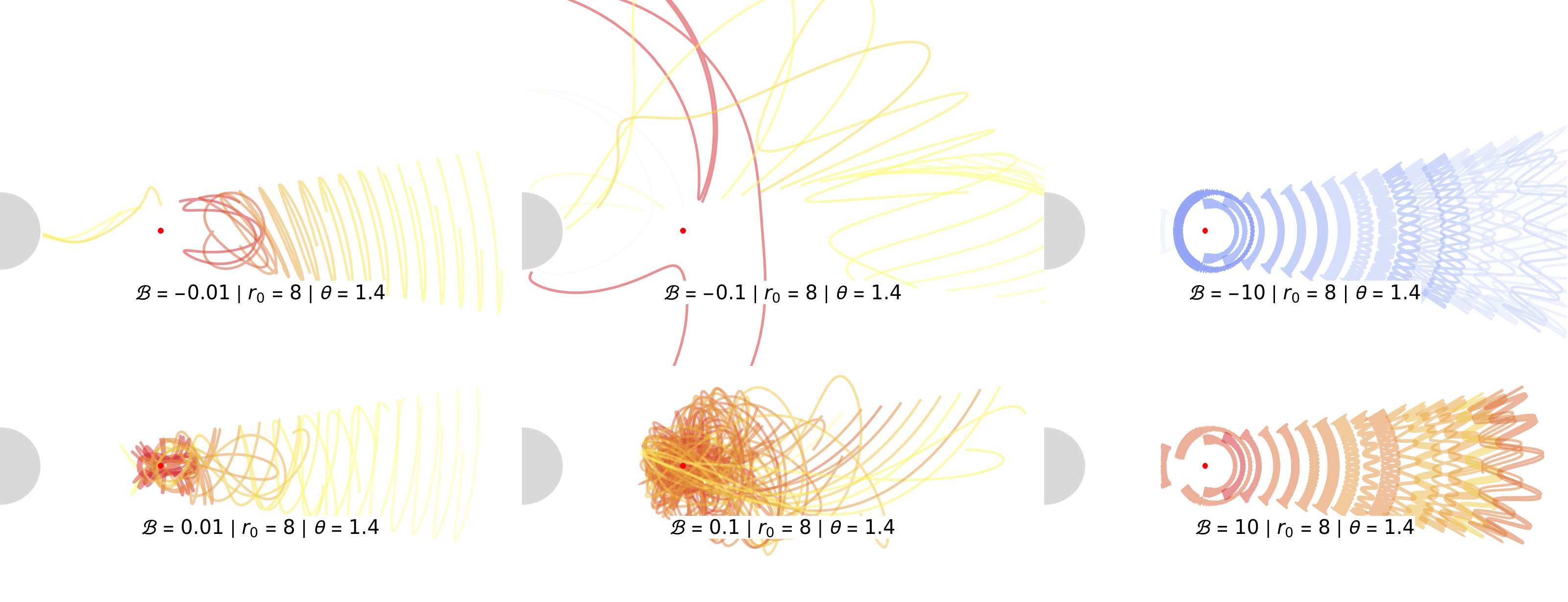}
\caption{Visualization of the probability density distribution of charged particles originating from an ionized Keplerian disk, see Fig.~\ref{fig:DISK}. The simulations consider various initial values of the magnetic field strength parameter $\cb$, associated with a current loop (marked by a red dot) located above neutral particle ISCO ($r_{\rm ISCO}=6$). 
Trajectories are color-coded according to the distance they travel in toroidal direction $\phi(\tau_{\rm END})$. Particles which have been moving around BH in counterclockwise direction ($\phi(\tau)>0$) will have colors from yellow (low $\phi(\tau_{\rm END})$) to orange (high $\phi(\tau_{\rm END})$), particles moving in clockwise direction ($\phi(\tau)<0$) will have colors from light blue to darker blue. 
%
For the weak magnetic field strengths and attractive Lorentz force ($\cb = 0.01$ and $\cb = 0.1$) the ionized particles tend to accumulate near the location of the current loop, for repulsive Lorentz force ($\cb = -0.01$ and $\cb = -0.1$) particles are trying to avoid the current loop location. For stronger magnetic field strengths ($\cb =\pm 10$), the magnetic influence dominates the dynamics, particles are expelled from the current loop vicinity, constrained to follow magnetic field lines, exhibiting mirror motion in the $\theta$-direction and azimuthal drift around the central BH.
\label{fig:SWARM1}
}
\end{figure*}

\begin{figure*}
\centering
\includegraphics[width=\textwidth]{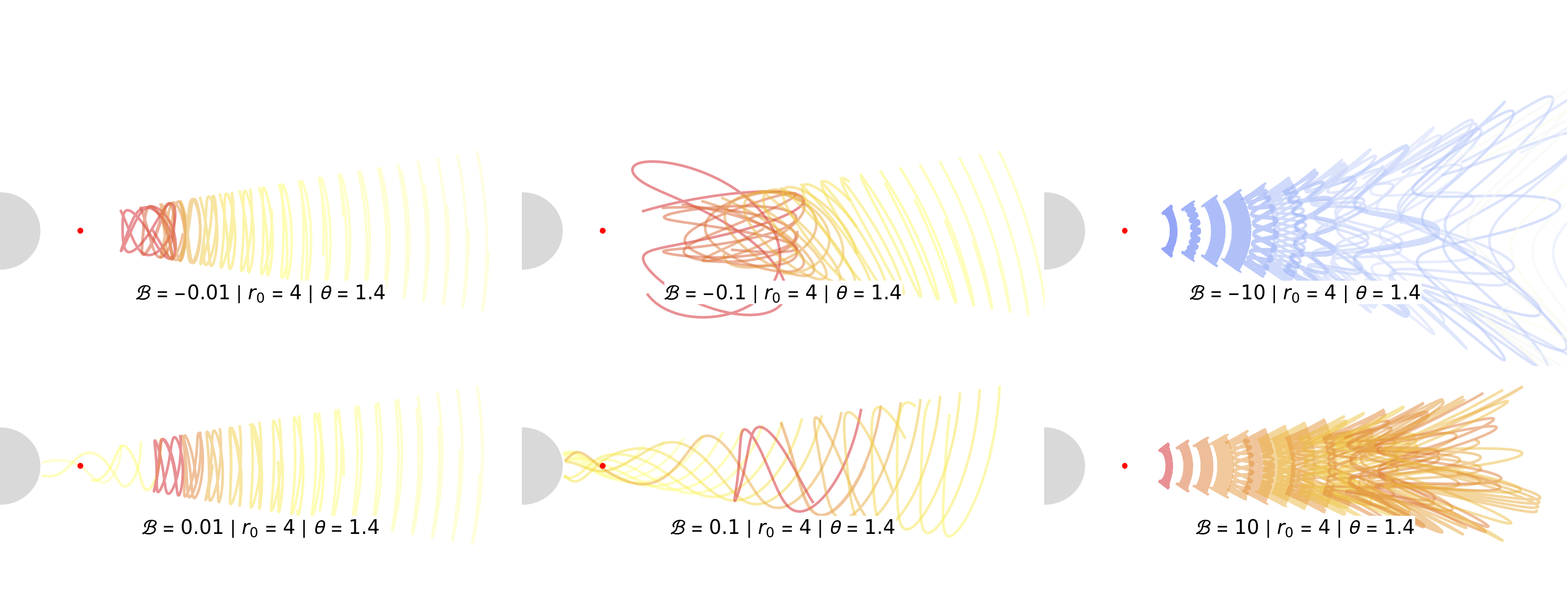}
\caption{Visualization of the probability density distribution of charged particle trajectories originating from an ionized Keplerian dis as presented in Fig.~\ref{fig:SWARM1}, but now for current loop located below neutral particle ISCO. Now the presence of loop magnetic filed will helps to the particle to settle inside.
\label{fig:SWARM2}
}
\end{figure*}

\section{Current ring stability}

When gravitational interaction is the dominant force, the equilibrium configuration is typically spherical, as exemplified by stars and BHs. In contrast, electromagnetic interaction, due to the coexistence of both attractive and repulsive forces, leads to qualitatively different structures. In plasma, matter preferentially organizes into elongated, filamentary configurations, often twisted into helices or forming even more complex morphologies. Such filaments are usually confined by their self–generated magnetic fields; for example, helical filaments, characterized by magnetic fields with both azimuthal and axial components, can achieve long–term stability. An additional effect, plasma diamagnetism, emerges from the gyromotion of charged particles around magnetic field lines, whereby each particle induces a magnetic field oriented oppositely to the external field. In this section, we try to address the problem of stability of the current-loop-generated BH magnetosphere using the dynamics of charged test particles.

To populate the BH surrounding with charged particles we will assume quasi-neutral plasma orbiting the central BH along circular on many quasi-circular geodesics forming a thin Keplerian disk as original source. The ionized Keplerian disks were studied for the non-rotating BHs \cite{Kol-Stu-Tur:2015:CLAQG:,Pan-Kol-Stu:2019:EPJC:,Kol-Sha-Tur:2023:EPJC:,Ken-etal:2024:PRD:} and for the rotating BHs in \cite{Stu-Kol:2016:EPJC:,Kop-Kar:2018:APJ:,Kop-Kar:2020:APJ:}. The test particle mass and mechanical momenta before (\mJ) and after the ionization / escape from the accretion flow (\mD) are conserved 
\beq 
m_{(\mJ)} = m_{(\mD)}, \quad p_{(\mJ)}^{\mu} = p_{(\mD)}^{\mu}. \label{ioniz}
\eeq
The velocity of the particle has not changed during ionization (\ref{ioniz}), but the motion constants, given by generalized (canonical) momenta (\ref{particleMOM}), will now be changed by the electromagnetic field presence. The test particle mass and mechanical momenta remain the same during ionization (time $\tau_0$) and hence we can simply obtain the initial four-velocity of the charged particle $u^{\mu}_{(\mD)}(\tau_0)$ from the neutral particle four-velocity $u^{\mu}_{(\mJ)}(\tau_0)$ -- the particle speed is not changing during ionization, but after the ionization, the charged particle starts to feel the LF determined by the MF. 

Drift motions of individual charged particles in magnetic and/or gravitational fields can raise macroscopic manifestations of currents in space plasma. In Fig.~\ref{fig:DISK} we see accumulation of ionized particles around the current loop from the originally neutral thin accretion disk. To see if the charged particles orbiting attracted by the current loop magnetic field will give rise to cumulative macroscopic current, we color the individual particle trajectory in Figs.~\ref{fig:SWARM1},\ref{fig:SWARM2} according to their displacement in the toroidal $\phi$ direction. For positive (counterclockwise) motion, red is used, while for negative (clockwise) motion, blue is applied.

In Fig.~\ref{fig:SWARM1}, we see that when the current loop is located above the edge of the neutral thin disk ($r_0 > r_{\rm ISCO} = 6$), an accumulation of charged particles around the current loop occurs only for weak magnetic field strengths ($\cb = 0.01$ and $\cb = 0.1$) under the action of an attractive Lorentz force. In this regime, the accumulated particles move in the positive toroidal direction (yellow to red colors), and their collective motion generates a ring current opposing the current loop, thereby reducing the original magnetic field. When the Lorentz force is repulsive ($\cb = -0.01$ and $\cb = -0.1$), the particles are still moving counterclockwise, but they are expelled from the vicinity of the loop. In this case, their collective motion would produce a ring current aligned with the original current loop; however, due to the absence of significant particle accumulation, the effect on the magnetic field remains negligible.

For large values of the magnetic parameter ($|\cb| = 10$, strong magnetic field), particles follow magnetic field lines and tend to avoid the intense field near the current loop. In both cases, Lorentz attractive ($\cb=10$) and Lorentz repulsive ($\cb=-10$), the currents generated by the ionized particle motion reduce the magnetic field originally produced by the current loop within the Keplerian disk.

In Fig.~\ref{fig:SWARM2}, we consider the scenario in which the current loop is located below the edge of the neutral thin disk. No significant accumulation of particles is observed; instead, ionized particles overflow from the Keplerian disk toward the central BH. For strong Lorentz forces ($\cb = \pm 10$), charged particles again avoid regions of intense magnetic field near the loop. They follow Larmor orbits around magnetic field lines, undergoing mirror motion in the vertical direction while simultaneously drifting azimuthally around the central object. In this regime, charged particles exhibit a screening response: Their collective dynamics generate a ring current that effectively reduces the magnetic field originally produced by the current loop. Similarly to the case of Earth’s magnetosphere, charged particles in a black hole magnetosphere sourced by a current loop attenuate the original magnetic field through their collective drift motion in all scenarios explored.

\section{Summary \& conclusion \label{sec:Summary}}


We have investigated the magnetic field generated by a toroidal current loop located in the equatorial plane of a non-rotating Schwarzschild BH by analyzing the dynamics of charged particles. This study is purely theoretical; however, the results can be applied to various astrophysical phenomena, such as charged particle acceleration~\cite{Tur-etal:2019:ApJ:} or quasiperiodic oscillations~\cite{Kol-Tur-Stu:2017:EPJC:}. Our primary objective was to explore the structure of the magnetic field generated by a toroidal current using charged particle dynamics. We examined the full general relativistic analytic solution for the magnetic field produced by the current loop, as given by Eq.~(\ref{eq:Aphi}) and derived in~\cite{Kof-Kot:2022:PRD:}. Furthermore, we established connections with the flat spacetime solution given by Eq.~(\ref{eq:Aflat}) from~\cite{Jackson:1998:book:}, as well as with the multipole expansion presented in Eq.~(\ref{eq:petterson}) from~\cite{Petterson:1975:PRD:}.


We present a theoretical analysis of charged particle motion in the exact solution of Maxwell's equations in the spacetime of a Schwarzschild BH. Below the current loop ($r<r_0$), the particle dynamics resemble motion in a uniform magnetic field \cite{Kol-Stu-Tur:2015:CLAQG:}, whereas outside the loop ($r>r_0$), the particle moves in a field similar to a magnetic dipole \cite{Vrb-Kol-Stu:2024:EPJC:}. The dynamics of charged particles can become highly non-linear and even chaotic when the Lorentz force is comparable in magnitude to the gravitational attraction of the BH. However, in the regime of strong magnetic fields ($|\cb|\gg{}1$), the motion of charged particles becomes adiabatic, and the particles closely follow magnetic field lines. We numerically investigated the trajectories of charged particles and classified them on the basis of the orientation of the Lorentz force relative to the current loop. Two distinct configurations emerge: attractive ($\cb>0$), where the Lorentz force is directed toward the loop, and repulsive ($\cb<0$), where the force points away. In the attractive case, the effective potential $V_{\rm eff}(r,\theta)$ develops a depression near the current loop, where particles can become trapped. Due to the divergence of the vector potential $A_\phi(r,\theta)$ in the analytic solution (\ref{eq:Aphi}) at the location of the current loop ($r=r_0,\theta=\pi/2$), the effective potential $V_{\rm eff}(r,\theta)$ also exhibits a divergence at this radius. As a result, motion of particles is forbidden precisely at the loop location, although orbits around it are still possible. From a physical perspective, the infinitesimally thin current loop and the associated infinitely high barrier in $V_{\rm eff}(r,\theta)$ are idealized and unrealistic. A more realistic configuration would involve a current loop with finite thickness, which could allow the motion of charged particles even within its interior.

We investigated the existence and stability of circular orbits of charged particles in current loop BH magnetosphere. For high values of the magnetic parameter $\cb$, stable circular orbits exist even in close proximity to the event horizon of the BH. In contrast, for lower values of $\cb$, the situation becomes more complex, and there exist regions where circular orbits of charged particles are not allowed. Stable circular orbits correspond to the minima of the effective potential $V_{\rm eff}(r,\theta)$ and represent the energy minima for bound (non-circular) orbits around the central object. Ensembles of ionized particles on bound orbits near the minimum of $V_{\rm eff}(r,\theta)$ may form radiation belts in the inner magnetosphere of the BH, provided the magnetic field $|\cb|$ is sufficiently strong. In the BH magnetosphere, minima of $V_{\rm eff}(r,\theta)$ may occur both in the equatorial plane and off the equatorial plane. The existence of off-equatorial minima, and the corresponding off-equatorial structures composed of trapped charged particles, is forbidden in the case of a uniform magnetic field around a non-rotating Schwarzschild BH, but becomes possible in the presence of a dipolar magnetic field or for a rotating Kerr BH. The off equatorial circular orbits are allowed only above the current loop radius $r>r_0$ and only for Lorentz repulsive case $\cb<0$.
The existence, shape, and spatial distribution of such radiation belts can be significantly influenced by electromagnetic waves propagating through the BH magnetosphere, suggesting a promising direction for future research.

In classical electrodynamics~\cite{Jackson:1998:book:}, free charged particles moving in an external magnetic field tend to weaken this field by generating their own magnetic field oriented in the opposite direction. When the motion of charged particles is governed by a combination of the Lorentz force and gravitational attraction, as is the case in the Earth's magnetosphere, a ring current can form due to charged particles drift in the radiation belts, which likewise acts to reduce the external magnetic field. We have demonstrated that free charged particles are expelled from the vicinity of the current loop radius due to the strong magnetic field near the loop, although particles can still follow bounded orbits encircling the current loop. In the case of an attractive Lorentz force ($\cb>0$), we observe an accumulation of charged particles near the loop, and the resulting collective current generated by these particles acts in opposition to the original current loop. It is evident that these effects are not exclusive to curved spacetime and would also arise in flat spacetime configurations. However, one of the most significant differences introduced by general relativity is the existence of ISCO for charged particles, which may serve as a lower bound on the location of stable radiation belts in the BH magnetosphere.


We have chosen a basic approach to the BH magnetosphere modeling using analytical solution of axially symmetric current, which could serve as just one element of more complete current distribution which can be obtained by summation of many current loops due to linearity of Maxwell equations. In the Force-Free Electrodynamics (FFE) approach or the GRMHD approach, which are both nonlinear, we can not just add two solutions and create a new one. Moreover, there are only a limited number of analytical solutions known in the FFE and GRMHD approximation, and we are relying on numerical simulations. For example, at the present time, there is no known full GRMHD analytical solution with poloidal magnetic field. For the initial conditions in the GRMHD simulations, a fluid hydrodynamical thick tori \cite{Koz-Jar-Abr:AAP:1978:} with artificially superimposed magnetic configuration is used, hoping to evolve to the full GRMHD solution during the simulations. Mostly a magnetic field following the contours of the mass density has been used, but in \cite{Sap-Jan:2019:APJ:,Jan-Jam-Pal:2021:APJ:} the magnetic field of a current loop in flat spacetime (\ref{eq:Aflat}) has been used. The complete GR solution (\ref{eq:Aphi}) introduced in \cite{Kof-Kot:2022:PRD:} will be a much better choice for such very thin torus. 

In realistic situations, the Dirac delta current loop around a black hole (\ref{eq:Aphi}) will likely be replaced by some form of current density associated with the orbiting mass distribution. Moreover, there is the problem of the four-potential diverging exactly at the current radius. Such problems can be avoided if we truncate the four potentials generated by the current loop at some value and fill the restricted toroidal region with a current density distribution, so that the Maxwell equations are satisfied. This substituted current must be floating in the toroidal direction at the boundary of the section, assuming the interior is empty. Another approach to a more realistic current distribution is to start with a heuristic model for the magnetic field, such as (\ref{power1st}), and then compute from Maxwell’s equations the current required to sustain it. The problem of reconciling the current loop solution with the matter distribution that generates it could provide insight into how to construct a complete analytic GRMHD solution.


Charged particles moving near a toroidal current loop help us better understand the formation of toroidal plasma structures, or plasma pinches, observed in GRPIC simulations of collisionless plasmas \cite{Bra-Rip-Phi:2021:PRL:,Mer-Cam-Rez:2025:}. In these simulations, magnetic fields naturally form around toroidal currents, resembling those around current loops, and tend to trap and guide charged particles. 
This work could be useful as an analytic model for a current-induced magnetic reconnection model, which is essential for high-energy processes near BH \cite{Bra-Rip-Phi:2021:PRL:,Par-Phi-Cer:2019:PRL:,Mer-Cam-Rez:2025:}. A well-constructed magnetic field model determines the location of current sheets and magnetic null points, regions where the magnetic field is central to particle acceleration and energy dissipation. 
Future studies should focus on incorporating finite-thickness current distributions and investigating their role in self-consistent field-particle interactions in full 3D relativistic plasma models.






\begin{acknowledgments}
This work was partially supported by the Czech Science Foundation Grant (GA\v{C}R) No.~\mbox{23-07043S}. 
\end{acknowledgments}


\providecommand{\noopsort}[1]{}\providecommand{\singleletter}[1]{#1}%

\end{document}